\documentclass[conference,compsoc]{IEEEtran}
\usepackage[T1]{fontenc}
\usepackage{amsmath}
\usepackage{amssymb}
\usepackage[utf8]{inputenc}
\usepackage{graphicx}
\usepackage{booktabs}
\usepackage{paralist}
\usepackage[table]{xcolor}
\usepackage[caption=false]{subfig}
\usepackage{enumitem}
\usepackage{multirow}
\usepackage{tabularx}
\usepackage{array}
\usepackage{listings}
\usepackage{cite}
\usepackage{flushend}
\usepackage{balance}
\usepackage{mdframed}
\usepackage{float}
\usepackage{placeins}

\usepackage{tikz}
\usetikzlibrary{arrows.meta, positioning, calc, backgrounds, fit}

\DeclareUnicodeCharacter{03A3}{$\Sigma$}    
\DeclareUnicodeCharacter{03C6}{$\phi$}      
\DeclareUnicodeCharacter{00B5}{$\mu$}        
\DeclareUnicodeCharacter{2191}{$\uparrow$}   
\DeclareUnicodeCharacter{2193}{$\downarrow$} 
\DeclareUnicodeCharacter{00D7}{$\times$}     

\newcommand{\df}{\mathit{df}}

\begin{document}

\title{NLLog: Lightweight, Explainable SOC Anomaly Detection\\via Log-to-Language Rewriting}



\author{
  \IEEEauthorblockN{
    Samuel Ndichu\textsuperscript{1},
    Tao Ban\textsuperscript{1},
    Seiichi Ozawa\textsuperscript{2},
    Takeshi Takahashi\textsuperscript{1},
    Daisuke Inoue\textsuperscript{1}
  }
  \IEEEauthorblockA{
    \textsuperscript{1}National Institute of Information and Communications Technology, Tokyo, Japan\\
    \textsuperscript{2}Kobe University, Kobe, Japan\\
    \{ndichu, bantao, takeshi\_takahashi, dai\}@nict.go.jp,\;
    ozawasei@kobe-u.ac.jp
  }
}

\maketitle

\begin{abstract}
System-generated logs underpin security monitoring, yet their rigid template-based format hinders both automated analysis and human comprehension.
We present \textbf{NLLog (Natural-Language Log)}, a lightweight pipeline that deterministically rewrites parsed templates into WHO--WHAT--SEVERITY sentences, pools them with term-frequency-inverse-document-frequency weighting, classifies sessions with tree ensembles, and back-projects evidence with TreeSHAP for analyst review.
On Hadoop Distributed File System (HDFS) and Blue Gene/L (BGL) corpora, NLLog exceeds two reproduced matched-protocol baselines; across HDFS, BGL, and the AIT Alert Data Set, it sustains low false-positive rates with commodity-hardware latency suitable for security operations center triage.
Coverage, sparse-versus-dense, faithfulness, and adversarial ablations show that fallback sufficiency is corpus-dependent, that an enrollment-time coverage check can surface refinement requirements before deployment, and that an auditable deterministic rewrite combined with lightweight dense encoding provides a measurable representation layer for log-anomaly detection and triage.
\end{abstract}


\sloppy

\section{Introduction}
\label{sec:introduction}

Security operations centers (SOCs) serve as the first line of defense for modern enterprises.  
SOC analysts investigate millions of log entries to detect and respond to threats. This persistent \emph{alert fatigue problem}, the difficulty of isolating genuine threats from an overwhelming number of machine-generated warnings, has endured for decades~\cite{axelsson2000base,sommer2010outside,sharma2021survey}. When critical alerts are drowned out by low-value noise, attacker dwell time increases and organizational risk escalates.

\textbf{NLLog} (\emph{Natural-Language Log}) targets one narrow part of this broader problem: the semantic opacity of template-based logs. Prior log anomaly detection work spans template heuristics, hand-engineered statistical features, and deep or language-model-based methods~\cite{he2017drain,he2016experience,du2017deeplog,zhu2022survey,lou2010invariantmining,meng2020survey,zhang2019robust,nedelkoski2020selfsupervised,pang2021deep,yang2025logllama}, including semi-supervised parsing pipelines such as PLELog~\cite{yang2021plelog}, no-parsing semantic models such as NeuralLog~\cite{le2021neurallog}, and large language model (LLM)-based approaches such as LogPrompt~\cite{10554918} and LogGPT~\cite{10386543}. These approaches can be effective, but they can also require more manual effort, computation, or tuning than lightweight SOC deployments allow. Pre-trained language models (LMs) offer useful semantics without fine-tuning~\cite{devlin2018bert,wang2020minilm}, yet raw logs contain many non-linguistic tokens that make direct encoding difficult~\cite{le2022log,yang2021semi,le2021neurallog,guo2021logbert}.

Instead of adapting larger models to raw logs, NLLog inserts a deterministic rewriting layer in front of a frozen encoder. For template-based software logs, NLLog parses each line with Drain3~\cite{he2017drain} and converts the resulting template into an analyst-readable WHO--WHAT--SEVERITY (WWS) sentence. For intrusion detection system (IDS) alert records such as AIT-ADS, where alerts are already structured textual records, NLLog uses the normalized alert signature as an alert-name WWS analogue. The resulting text is then embedded with a frozen sentence encoder, aggregated via term-frequency-inverse-document-frequency (TF--IDF) weighting~\cite{salton1988term}, and classified by a tree ensemble whose decisions are back-projected to the most influential sentences via TreeSHAP (Tree SHapley Additive exPlanations). The rewriting step is deterministic normalization rather than prompt engineering, which keeps the system reproducibly CPU-deployable.

We evaluate NLLog on three public datasets spanning clusters, supercomputers, and real-world alerts: Hadoop Distributed File System (HDFS)~\cite{xu2009hdfs,zhu2023loghub}, Blue Gene/L supercomputer logs (BGL)~\cite{oliner2007supercomputers,zhu2023loghub}, and the AIT Alert Data Set (AIT-ADS)~\cite{10.1145/3675741.3675748,9866880}.  
On HDFS and BGL under our matched protocol, NLLog achieves higher scores than two reproduced baselines (DeepLog and LogBERT); we also report published results from recent systems as reference points rather than direct head-to-head comparisons. On AIT-ADS, NLLog maintains high precision and low false-positive rates on a more operational alert corpus. In addition, NLLog generates human-readable explanations intended to support analyst triage.

NLLog is complementary to provenance-graph intrusion detection systems such as ORTHRUS, PROGRAPHER, and ThreatRace~\cite{orthrus,prographer,threatrace}: those systems target richer audit telemetry and system-wide attack tracing with graph-construction overhead and heavier instrumentation, whereas NLLog targets sessionized, line-oriented logs available in lightweight SOC pipelines. A protocol-matched comparison would require either adapting NLLog to system-call traces or evaluating provenance systems on a line-log corpus, which is outside the scope of the current evaluation.

\noindent \textbf{Contributions:} This work makes the following key contributions:
\begin{itemize}
    \item \textbf{Deterministic language-aligned rewriting.} NLLog shows that deterministic WWS rewriting can make line-oriented logs more model-compatible and analyst-readable without fine-tuning or prompting; Section~\ref{sec:rq1} reports an 8.47~pp F\textsubscript{1} gain on BGL over raw template inputs (HDFS is near-ceiling for all variants), and a coverage and fallback-only ablation (Table~\ref{tab:wws_coverage}) quantifies when generic fallback rewriting is sufficient and when deterministic corpus-specific lexical refinement is needed.
    \item \textbf{Lightweight SOC deployment path.} Frozen MiniLM, TF--IDF pooling, and tree classifiers achieve competitive matched-protocol results while remaining CPU-deployable at ${\sim}10\,\text{ms}$/session with no GPU requirement.
    \item \textbf{Attribution as structured triage evidence.} TreeSHAP back-projection returns ranked WWS sentences rather than causal root-cause explanations; faithfulness and mimicry-padding tests quantify when the evidence remains decision-relevant and where dilution begins to degrade it. Analyst-perceived utility remains to be validated in a controlled study.
    \item \textbf{Conservative evaluation framing.} Reproduced baselines are separated from published reference points, and AIT-ADS chronological and sessionization checks extend evaluation beyond widely used HDFS/BGL benchmark scores.
\end{itemize}

\section{Background and Threat Model}
\label{sec:background}

This section frames the deployment context that NLLog targets and the threat model under which it is evaluated. We first describe the operational SOC pipeline NLLog plugs into, then state the adversary's capabilities, the attack classes in scope, and what NLLog does not defend against.

\subsection{Operational Context and System Design}
\label{sec:operational}
Enterprise systems emit millions of timestamped events per day. Each log line adheres to a rigid template in which runtime variables (e.g., IP addresses, block IDs) are substituted, producing terse, cryptic messages that lack syntactic structure. SOC analysts face three compounding challenges: \textbf{volume}, \textbf{heterogeneity}, and \textbf{semantic opacity}. NLLog addresses the last by deterministically generating a self-contained WWS sentence for each template (Table~\ref{tab:wws-examples}), preserving core semantics while masking high-cardinality tokens. The framework inserts a lightweight preprocessing layer into existing security information and event management (SIEM) pipelines (Fig.~\ref{fig:system_framework}) and is intended to sit between SIEM output and Level-1 analyst triage.
\subsection{Threat Model}
\label{sec:threat}
\textbf{Goal:} The adversary seeks to perform malicious actions (e.g., lateral movement, privilege escalation, data exfiltration) while avoiding detection by NLLog.

\textbf{Capabilities:} The attacker may generate logs via normal system APIs and manipulate free-variable values (e.g., IPs, filenames). We assume that the logging infrastructure's integrity is maintained: the attacker does not alter hard-coded template strings embedded in binaries, disable logging, modify NLLog binaries or model weights, or interfere with its execution environment. This bounds our focus to adversaries who operate through legitimate execution paths.
Section~\ref{sec:discussion} discusses mitigations when this assumption is violated.

\textbf{Scope:} NLLog is a lightweight front-end SOC triage detector for environments with trusted log generation. We focus on \emph{semantic evasion attacks}, where adversaries introduce log events whose content deviates from learned benign distributions. Two classes of attack are explicitly out of scope. First, because NLLog uses TF--IDF weighted pooling (Section~\ref{sec:lms}), it detects deviations in event frequency and composition rather than temporal ordering; attacks whose anomalous signature lies purely in the sequence of individually benign events (for example, some lateral-movement and living-off-the-land patterns) require a complementary sequence-aware detector. Second, NLLog does not model cross-session or cross-host correlations. NLLog is intended to run alongside such detectors in a layered SOC architecture. We empirically characterize long-horizon retraining-time contamination and deliberate IDF poisoning (Section~\ref{sec:adv-eval}, Appendix~\ref{app:idf_poisoning}), without claiming full robustness; IDF weights are computed on the training split and frozen at test time. Mitigations against template-manipulating insider threats (template integrity attestation, cross-host correlation) are discussed in Section~\ref{sec:discussion}.

\section{Methodology}
\label{sec:methodology}

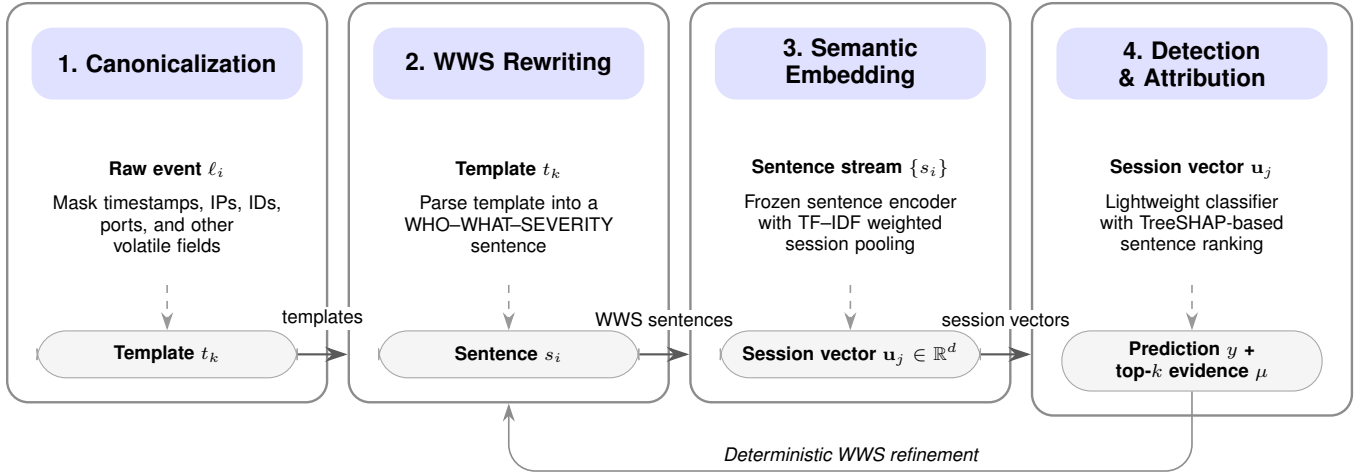
\begin{figure*}[htbp]
\centering
\begin{tikzpicture}[
    font=\sffamily,
    phase/.style={font=\footnotesize\scshape\bfseries, text=black!70},
    module/.style={draw=black!45, line width=0.9pt, rounded corners=8pt, fill=white},
    stagehead/.style={draw=none, fill=blue!12, rounded corners=8pt, text width=3.35cm, minimum height=0.95cm, align=center, font=\small\bfseries\sffamily},
    stagebody/.style={draw=none, text width=3.05cm, minimum height=2.05cm, align=center, font=\scriptsize\sffamily, inner sep=0pt},
    artifact/.style={draw=black!35, fill=black!4, rounded corners=11pt, text width=2.95cm, minimum height=0.62cm, align=center, inner xsep=7pt, inner ysep=4pt, font=\scriptsize\sffamily\bfseries},
    flow/.style={-{Stealth[length=3mm, width=2mm]}, thick, draw=black!65},
    auxflow/.style={-{Stealth[length=2.4mm, width=1.6mm]}, semithick, dashed, draw=black!40},
    feedback/.style={-{Stealth[length=2.4mm, width=1.6mm]}, semithick, draw=black!45, rounded corners=10pt}
]

    \node[stagehead] (h1) {1. Canonicalization};
    \node[stagebody, below=0.38cm of h1] (b1) {\textbf{Raw event} $\ell_i$\\[5pt]
        Mask timestamps, IPs, IDs,\\
        ports, and other\\
        volatile fields};
    \node[artifact, below=0.60cm of b1] (a1) {Template $t_k$};

    \node[stagehead, right=0.92cm of h1] (h2) {2. WWS Rewriting};
    \node[stagebody, below=0.38cm of h2] (b2) {\textbf{Template} $t_k$\\[5pt]
        Parse template into a\\
        WHO--WHAT--SEVERITY\\
        sentence};
    \node[artifact, below=0.60cm of b2] (a2) {Sentence $s_i$};

    \node[stagehead, right=0.92cm of h2] (h3) {3. Semantic Embedding};
    \node[stagebody, below=0.38cm of h3] (b3) {\textbf{Sentence stream} $\{s_i\}$\\[5pt]
        Frozen sentence encoder\\
        with TF--IDF weighted\\
        session pooling};
    \node[artifact, below=0.60cm of b3] (a3) {Session vector $\mathbf{u}_j \in \mathbb{R}^d$};

    \node[stagehead, right=0.92cm of h3] (h4) {4. Detection \& Attribution};
    \node[stagebody, below=0.38cm of h4] (b4) {\textbf{Session vector} $\mathbf{u}_j$\\[5pt]
        Lightweight classifier\\
        with TreeSHAP-based\\
        sentence ranking};
    \node[artifact, below=0.60cm of b4] (a4) {Prediction $y$ + top-$k$ evidence $\mu$};

    \begin{scope}[on background layer]
        \node[module, fit=(h1) (b1) (a1), inner sep=9pt] (m1) {};
        \node[module, fit=(h2) (b2) (a2), inner sep=9pt] (m2) {};
        \node[module, fit=(h3) (b3) (a3), inner sep=9pt] (m3) {};
        \node[module, fit=(h4) (b4) (a4), inner sep=9pt] (m4) {};
    \end{scope}

    \draw[flow] (a1.east) -- node[midway, above=9pt, fill=white, inner sep=1pt, font=\scriptsize\sffamily] {templates} (m2.west |- a1.east);
    \draw[flow] (a2.east) -- node[midway, above=9pt, fill=white, inner sep=1pt, font=\scriptsize\sffamily] {WWS sentences} (m3.west |- a2.east);
    \draw[flow] (a3.east) -- node[midway, above=9pt, fill=white, inner sep=1pt, font=\scriptsize\sffamily] {session vectors} (m4.west |- a3.east);

    \draw[auxflow] (b1.south) -- (a1.north);
    \draw[auxflow] (b2.south) -- (a2.north);
    \draw[auxflow] (b3.south) -- (a3.north);
    \draw[auxflow] (b4.south) -- (a4.north);

    \draw[feedback] (a4.south) -- ++(0,-1.05) -| (m2.south)
        node[pos=0.25, above, font=\scriptsize\itshape\sffamily] {Deterministic WWS refinement};

    \coordinate (semrep) at ($(m2.north)!0.5!(m3.north)$);
    \node[phase] at ($(m1.north)+(0,0.72)$) {Log Distillation};
    \node[phase] at ($(semrep)+(0,0.72)$) {Semantic Representation};
    \node[phase] at ($(m4.north)+(0,0.72)$) {Detection and XAI};

\end{tikzpicture}
\caption{NLLog pipeline from canonicalized logs to anomaly predictions and top-$k$ sentence evidence.}
\label{fig:system_framework}
\end{figure*}

NLLog comprises four stages (Fig.~\ref{fig:system_framework}): (1) canonicalize raw logs, (2) parse templates and rewrite each event as a WHO--WHAT--SEVERITY (WWS) sentence, (3) embed and TF--IDF pool sentences into session vectors, and (4) classify sessions and attribute decisions back to top-$k$ sentences using TreeSHAP.
\subsection{Raw Log Canonicalization}
\label{sec:preprocess}

Raw production logs contain volatile fields (e.g., timestamps, IPs, ports, block identifiers, and numeric offsets) that inflate the vocabulary and obscure semantic structure~\cite{he2016experience,zhu2022survey}. NLLog transforms each log line into a canonicalized form $\widetilde{m}_i$\,that preserves core semantics while masking these fields; Appendix~\ref{app:canonicalization} summarizes the substitutions.

\subsection{Natural-Language Log Generation}
\label{sec:nllog}

\paragraph{Motivation for the WWS design choice}
Raw Drain3 template identifiers are compact but semantically opaque: a token such as \texttt{E23} carries no information about actor, action, or severity without a separate lookup table, and pre-trained sentence encoders were never exposed to this vocabulary during pre-training. The WHO--WHAT--SEVERITY (WWS) rewrite trades that opacity for an analyst-readable surface form, e.g., \emph{``DataNode received block \textlangle{}BLOCK\_ID\textrangle{} from \textlangle{}IP\textrangle{}:\textlangle{}PORT\textrangle{} (info)''}, while remaining fully deterministic and requiring no model fine-tuning or prompt construction. This normalization is central to NLLog for three reasons: (1) it puts log vocabulary into a natural-language-like form better suited to frozen sentence encoders; (2) it produces the self-describing sentences that TreeSHAP later returns as triage evidence; and (3) its fallback behavior is operationally measurable before deployment (Table~\ref{tab:wws_coverage}).

Each canonicalized log $\widetilde{m}_i$\,is transformed into a stylized natural-language sentence $s_i$ using a WWS schema. Structurally similar log lines are grouped into canonical event templates via Drain3~\cite{he2017drain}; Appendix~\ref{app:drain3} summarizes the parsing setup. We define the mapping
\[
s_i = \Phi(t_k, c_i, \text{lvl}_i) = \text{WHO}(c_i) \;\Vert\; \text{WHAT}(t_k) \;\Vert\; \text{SEV}(\text{lvl}_i),
\]
where $\Phi : \mathcal{T} \times \mathcal{C} \times \mathcal{L} \rightarrow \mathcal{S}$ maps each template--component--level triple to a sentence, making the input surface form more regular~\cite{meng2020survey,zhu2022survey}. The WWS rendering is deterministic once the template is available. The mappings are governed by three interpretable functions:
\begin{itemize}[leftmargin=*, itemsep=2pt]
  \item $\text{WHO}(c_i)$: component role (e.g., ``NameNode'', ``DataNode''), derived directly from the emitting component name.
  \item $\text{WHAT}(t_k)$: verb-object clause generated from template structure using simple heuristics. A small number of optional deterministic lexical refinements improve the readability of the WHAT clause; Table~\ref{tab:wws_coverage} isolates their contribution to detection accuracy via a fallback-only ablation. If the heuristic fails on a new or malformed template, NLLog falls back to wrapping the raw canonicalized string in a minimal WWS frame (WHO + raw template + SEV).
  \item $\text{SEV}(\text{lvl}_i)$: severity marker obtained from the log level.
\end{itemize}

As shown in Fig.~\ref{fig:system_framework}, the stylized sentence $s_i$ becomes the unit of semantic embedding; Table~\ref{tab:wws-examples} illustrates representative rewrites. Anchoring the rewrite on Drain3 templates keeps outputs reproducible and auditable; automating this layer while preserving those properties remains future work.

\begin{table*}[htbp]
  \centering
  \caption{Representative WHO--WHAT--SEVERITY (WWS) rewrites.}
  \label{tab:wws-examples}
  \renewcommand{\arraystretch}{1.15}
  \begin{tabular}{@{}lclp{6.5cm}@{}}
    \toprule
    {Dataset} & {Class} & {Raw log fragment} & {WWS sentence} \\ 
    \midrule
    \multirow{2}{*}{\textbf{BGL}}
        & Normal
        & \textsf{RAS IO INFO link-chip reconfigured successfully}
        & \textit{I/O Subsystem reconfigured a link-chip successfully (info).} \\
    & Anomaly
        & \textsf{RAS KERNEL FATAL rts panic! -- stopping execution}
        & \textit{Kernel runtime system encountered a panic and halted execution (fatal).} \\
    \multirow{2}{*}{\textbf{HDFS}}
        & Normal
        & \textsf{DataXceiver: Receiving block blk\_\dots src \dots dest \dots}
        & \textit{DataNode Data Receiver received block \textless{}BLOCK\_ID\textgreater{} from \textless{}IP\textgreater{}:\textless{}PORT\textgreater{} destined for \textless{}IP\textgreater{}:\textless{}PORT\textgreater{} (info).} \\
    & Anomaly
        & \textsf{Pending Replication Monitor: Block \dots replication timed out}
        & \textit{Replication Monitor reported a replication timeout for block \textless{}BLOCK\_ID\textgreater{} (warning).} \\
    \multirow{2}{*}{\textbf{AIT-ADS}}
        & Normal
        & \textsf{Wazuh: IDS event.}
        & \textit{Wazuh IDS logged an IDS event (info).} \\
    & Anomaly
        & \textsf{Suricata: Alert - ET INFO Observed DNS Query to .biz TLD}
        & \textit{Suricata IDS observed a DNS query to the {.biz} top-level domain (alert).} \\
    \bottomrule
  \end{tabular}
\end{table*}

\subsection{Semantic Embedding and Session Pooling}
\label{sec:lms}

The stylized sentences $s_i$ produced in Section~\ref{sec:nllog} are embedded into dense vectors via a lightweight language model. We adopt MiniLM-L6~\cite{wang2020minilm}, a distilled 6-layer transformer with 22M parameters, because it offers a favorable accuracy--latency trade-off for CPU deployment.

Each sentence $s_i$ is subword-tokenized and bracketed with special tokens:
\[
s_i \;\longmapsto\;
\langle\texttt{[CLS]},\,t_{i,1},\dots,t_{i,\ell_i},\,\texttt{[SEP]}\rangle.
\]
MiniLM produces contextual embeddings $\mathbf{h}_{i,j}\in\mathbb{R}^{384}$ at each token position $j$. To obtain a fixed-size vector, we apply mean pooling over all non-padding tokens:
\begin{equation}
\mathbf{v}_i
=\;
\frac{1}{\ell_i+2}\;
\sum_{j=0}^{\ell_i+1}\mathbf{h}_{i,j}
\in \mathbb{R}^{384}.
\label{eq:minilm-embed}
\end{equation}
We use mean pooling instead of the \texttt{[CLS]} embedding for its greater stability in embedding-based settings~\cite{reimers2019sbert}. 


\paragraph{Session-Level Aggregation}
To analyze complete sessions, we embed each log line individually, then aggregate the results. For session $j$, let:
\[
S_j = \{s_{j1}, \dots, s_{j n_j}\}
\]
denote its stylized sentence set.
For each stylized sentence \(s_{ji}\in S_j\), we write \(\mathbf{v}_{ji}\in\mathbb{R}^{384}\) for its MiniLM embedding obtained via mean pooling in equation~\eqref{eq:minilm-embed}. A naive average would treat all lines equally, inflating the impact of repetitive boilerplate logs. Instead, we apply TF--IDF weighting to prioritize rare, security-relevant messages.

Let $N = |\mathcal{S}|$ be the number of sentences in the corpus, $f_{t,s}$ the frequency of token $t$ in sentence $s$, and $\df(t)$ the number of sentences containing $t$. TF--IDF is computed as:
\[
\mathrm{tfidf}(t,s)
= f_{t,s} \cdot \log\!\Bigl(\tfrac{N}{1 + \df(t)}\Bigr).
\]
We then define a sentence-level weight~\cite{salton1988term}:
\begin{equation}
w_{ji}
= \frac{1}{|s_{ji}|}
  \sum_{t\,\in\,s_{ji}} \log\!\Bigl(\tfrac{N}{1 + \df(t)}\Bigr),
\label{eq:sent-weight}
\end{equation}
favoring semantically rare tokens that often signify error conditions or anomalies.
The final session vector is the TF--IDF-weighted average of its sentence embeddings:
\begin{equation}
\mathbf{u}_j
= \frac{\sum_{i=1}^{n_j} w_{ji} \,\mathbf{v}_{ji}}
       {\sum_{i=1}^{n_j} w_{ji} + \varepsilon},
\quad \varepsilon \approx 10^{-8}.
\label{eq:session-vector}
\end{equation}
This aggregation amplifies influential sentences while reducing noise from redundant status logs. By design, TF--IDF pooling discards temporal order: if two events are swapped chronologically, the resulting session vector $\mathbf{u}_j$ remains identical. This is an intentional trade-off that prioritizes rare-event detection over sequence awareness; Section~\ref{sec:discussion} discusses its consequences.
%
%
%
\subsection{Session Classification and Explainability}
\label{sec:inves}

After computing TF--IDF-weighted session embeddings $\mathbf{u}_j \in \mathbb{R}^{384}$ from Eq.~\eqref{eq:session-vector}, each session $j$ is associated with a binary label $y_j \in \{0,1\}$. We train tree-based classifiers, including Random Forest (RF)~\cite{breiman2001rf}, XGBoost (XGB)~\cite{chen2016xgboost}, and LightGBM (LGBM)~\cite{ke2017lightgbm}, to estimate the anomaly probability $\hat{p}_j = f(\mathbf{u}_j)$, followed by a threshold decision $\hat{y}_j = \mathbf{1}[\hat{p}_j \ge \tau]$, with $\tau = 0.5$ by default. Class imbalance is handled with standard class-balancing options and 3-fold cross-validated hyperparameter search. We report Precision, Recall, F\textsubscript{1}, false positive rate (FPR), and area under the precision-recall curve (AUC-PR).

We augment classification with a lightweight explainability layer that highlights the log lines within each session that most influenced the prediction, combining TF--IDF-weighted embeddings with TreeSHAP contribution scores.

\subsubsection{Feature Attribution via TreeSHAP}
TreeSHAP~\cite{lundberg2020local} computes exact, consistent Shapley values for tree-based models using the path-dependent algorithm, which runs in $\mathcal{O}(TLD^2)$ time for $T$ trees of maximum depth $D$ with $L$ leaves. For our LGBM configurations (300 trees, depth $\leq 6$), this yields sub-millisecond per-session computation. We use it to explain each session's anomaly score in log-odds space:
\begin{equation}
\operatorname{logit}(\hat{p}_j) = \phi_{j,0} + \sum_{k=1}^{384}\phi_{j,k},
\label{eq:treeshap-logit}
\end{equation}
where $\phi_{j,0}$ is the expected output and $\phi_{j,k}$ denotes the contribution of embedding dimension $k$ to session $j$'s prediction. This yields a feature attribution vector $\boldsymbol{\phi}_j \in \mathbb{R}^{384}$.

\subsubsection{Back-Projection to Log Lines}
To map these contributions back to individual log lines, we define a responsibility matrix $\mathbf{A}_j \in \mathbb{R}^{n_j \times 384}$:
\begin{equation}
A_{ji,k} = \frac{w_{ji}}{\sum_{i'} w_{ji'}} \cdot \frac{v_{ji,k}}{u_{j,k} + \varepsilon},
\end{equation}
%
%
where $v_{ji,k}$ is the $k$-th MiniLM embedding dimension for sentence $i$ in session $j$, and $w_{ji}$ is its TF--IDF weight. The small constant $\varepsilon$ stabilizes dimensions whose pooled value is close to zero. This quantity is used as a ranking-oriented attribution score rather than as an exact sentence-level Shapley decomposition.
We then compute the total contribution of sentence $i$ to the model's decision via:
\begin{equation}
\psi_{ji} = \sum_{k=1}^{384} A_{ji,k} \cdot \phi_{j,k}.
\label{eq:psi-ji}
\end{equation}
Positive $\psi_{ji}$ values indicate anomaly-promoting log lines and negative values normality-promoting lines, identifying the most anomaly-driving sentences at the line level. Because this projection is used only for ranking, large ratios near small pooled dimensions are treated as ranking signals rather than calibrated Shapley values.

\subsubsection{Template-Level Aggregation}
For higher-level insights, we aggregate $\psi_{ji}$ across lines that share the same template. Let $\text{tpl}(s_{ji}) = t$ be the Drain3 template of sentence $s_{ji}$. Then:
\begin{equation}
\Psi_{j,t} = \sum_{\{i:\,\text{tpl}(s_{ji}) = t\}} \psi_{ji},
\quad
\bar{\psi}_{j,t} = \frac{\Psi_{j,t}}{\#\{i:\,\text{tpl}(s_{ji}) = t\}}.
\label{eq:template-agg}
\end{equation}
Here, $\Psi_{j,t}$ captures the total contribution of template $t$, while $\bar{\psi}_{j,t}$ represents its per-occurrence average impact. We compute $\Psi_{j,t}$ and $\bar{\psi}_{j,t}$ at inference time and expose them through the released artifact for SOC-dashboard auditing, including optional per-class aggregation (TP, FP, TN, FN) that surfaces high-impact templates even when infrequent; the main-paper evaluation focuses on per-sentence evidence ($\psi_{ji}$), which is what the analyst sees. All explainability stages run on CPU within the deployment profile evaluated in this paper.

\section{Evaluation}
\label{sec:evaluation}

We evaluate NLLog along three axes: detection accuracy, generalizability across classifiers and data regimes, and operational robustness with explainability.

\subsection{Experimental Setup}
\label{sec:exp-setup}

All experiments were performed on a 32-core Intel Xeon Gold 6242 CPU @ 2.80\,GHz (64 threads, 22\,MB L3 cache), in a CPU-only environment with no GPU acceleration. Transformer inference used half precision (FP16) where supported by the local inference stack; otherwise FP32 was used.

\paragraph{End-to-End Latency}
We profiled the full NLLog pipeline on the same host; Table~\ref{tab:latency-breakdown} reports median and 95th-percentile (p95) cumulative latency per session.

\begin{table}[htbp]
\centering
\caption{Median and p95 end-to-end latency (ms / session).}
\label{tab:latency-breakdown}
\begin{tabular}{lcc}
\toprule
\textbf{Stage} & \textbf{Median} & \textbf{p95} \\ 
\midrule
Canonicalization           & 0.026 & 0.047 \\
WWS assembly               & 0.001 & 0.001 \\
MiniLM embedding           & 8.473 & 10.657 \\
TF--IDF aggregation         & 0.189 & 0.220 \\
\texttt{predict\_proba}    & 1.031 & 1.266 \\
TreeSHAP top-$k$           & 0.202 & 0.219 \\ 
\midrule
\textbf{End-to-end}        & \textbf{9.92} & \textbf{12.41} \\
\bottomrule
\end{tabular}

\vspace{0.5ex}
  {\footnotesize
    \textit{Note:} The p95 entry reports the latency below which 95\% of sessions complete.\par}
\end{table}

We evaluated six frozen encoders ranging from MiniLM-L6 (22M parameters) to BERT-large (335M), all with mean pooling~\cite{reimers2019sbert}. DeepLog~\cite{du2017deeplog} and LogBERT~\cite{guo2021logbert} are reproduced under our protocol; LogAnomaly~\cite{meng2019loganomaly}, LogLLM~\cite{guan2024logllm}, LogLLaMA~\cite{yang2025logllama}, and LogGPT~\cite{10386543} values are published results included as contextual reference points. DeepCASE~\cite{vanede2022deepcase} is reproduced only for the data-fraction analysis (Section~\ref{sec:rq2h6}).

We report Precision, Recall, F\textsubscript{1}, FPR, and AUC-PR (emphasizing AUC-PR on skewed datasets) over five random seeds (42, 1337, 31415, 2025, 8675309), using an 80:20 stratified split with 3-fold stratified cross-validation for model selection.

Table~\ref{tab:dataset_summary} summarizes the datasets. HDFS sessions are block-level traces; BGL uses 100-line non-overlapping windows~\cite{he2016experience}; AIT-ADS comprises 2.65M IDS alerts from Suricata, Wazuh, and AMiner across eight multi-stage attack scenarios, sessionized by source host and fixed-length time windows following the dataset's labeling scheme. For HDFS and BGL, WWS is a template-to-sentence rewrite; for AIT-ADS, NLLog uses the normalized alert signature as an alert-name WWS analogue, with richer non-leaky sparse baselines evaluated in Section~\ref{sec:rq2} (Tables~\ref{tab:sparse_dense} and~\ref{tab:ait_sparse_dense}).

\paragraph{Evaluation Hygiene}
All corpus-derived statistics (Drain3 templates, IDF weights, and deterministic lexical refinements) are computed on the training split and frozen at test time.

\begin{figure}[t]
\centering
\includegraphics[width=0.92\linewidth]{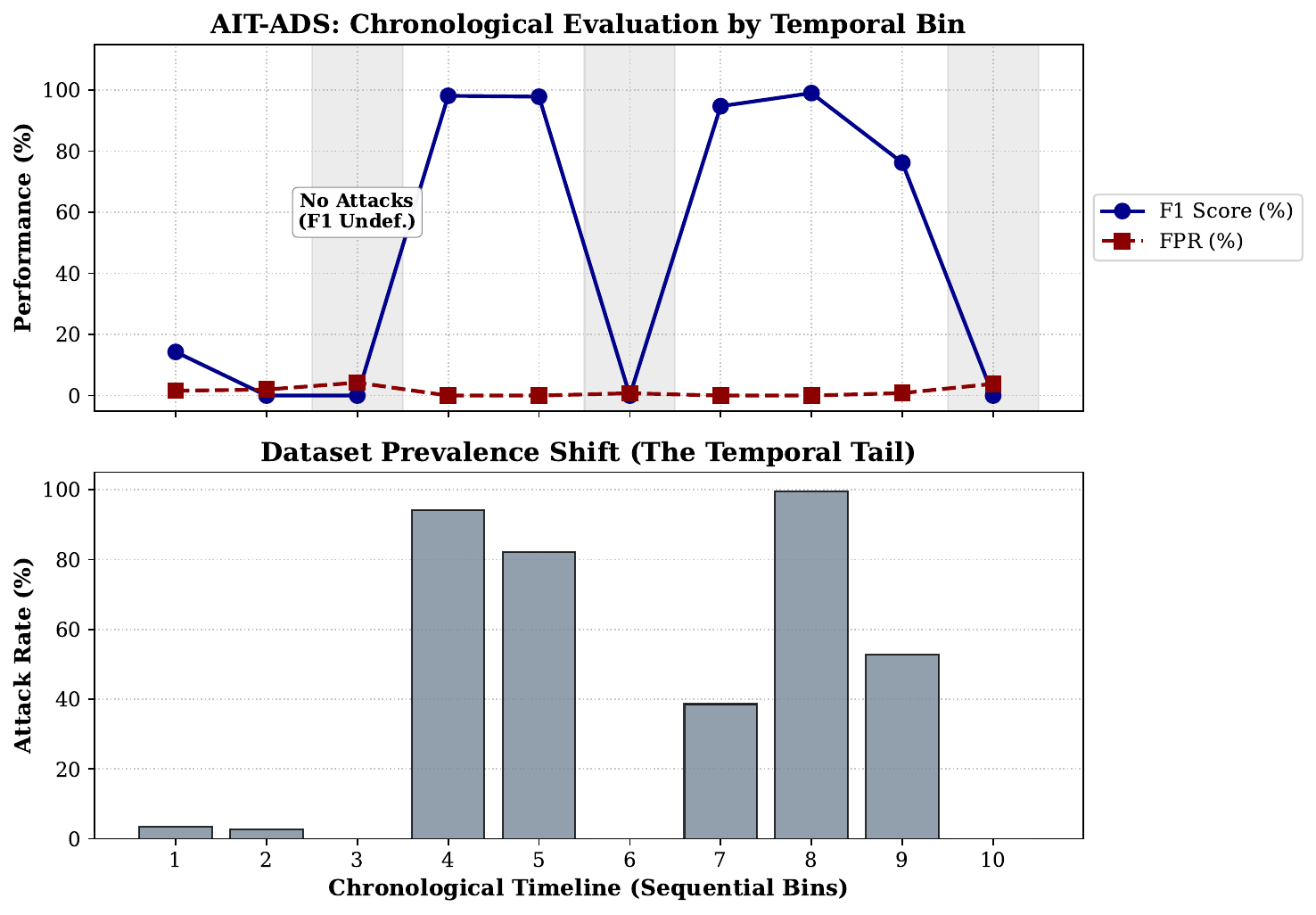}
\caption{AIT-ADS chronological evaluation.}
\label{fig:aitads_temporal}
\end{figure}

\paragraph{Chronological evaluation on AIT-ADS}
In addition to the 80/20 stratified split, we report a chronological evaluation on AIT-ADS (oldest 80\% train, newest 20\% test). LightGBM reaches 93.90\% F\textsubscript{1}, 98.40\% precision, 89.80\% recall, 0.90\% FPR, and 95.50\% AUC-PR, compared with 92.93\%, 99.18\%, 87.43\%, 0.14\%, and 92.38\% under the stratified split. Because the test tail has much higher anomaly prevalence (37.37\% vs.\ 10.51\% in training), this is a deployment-tail stress test rather than a pure temporal-drift measurement; Fig.~\ref{fig:aitads_temporal} summarizes the temporal behavior.

\paragraph{AIT-ADS sessionization sensitivity}
AIT-ADS test F\textsubscript{1} ranges from 80.00\% to 90.93\% across maximum session durations of 5--120 minutes (60 minutes best), with the 300\,s idle timeout fixed; Appendix~\ref{app:aitads_window} details the sweep.

\begin{table*}[htbp]
\centering
\caption{Summary of BGL, HDFS, and AIT-ADS dataset statistics.}
\label{tab:dataset_summary}
\begin{tabular}{lrrrrrrrr}
\toprule
\multirow{2}{*}{Dataset} & \multirow{2}{*}{\#Logs} & \multirow{2}{*}{\#Sessions} & \multicolumn{3}{c}{Training Data} & \multicolumn{3}{c}{Testing Data} \\
 &  &  & {\#Sess.} & {Anomaly} & {Imb(\%)} & {\#Sess.} & {Anomaly} & {Imb(\%)} \\
\midrule
BGL      & 4,747,963   & 47,479    & 37,983     & 3,842    & 10.12 & 9,496     & 960     & 10.11 \\
HDFS     & 11,175,629  & 575,061   & 460,048    & 13,470   & 2.93  & 115,013   & 3,368    & 2.93  \\
AIT-ADS  & 2,655,821   & 12,983    & 10,386     & 1,650      & 15.89  & 2,597     & 412       & 15.86  \\
\bottomrule
\end{tabular}

\vspace{0.5ex}
  {\footnotesize
  \textit{Note:} Imb(\%) is the percentage of anomalous sessions.\par}
\end{table*}

\subsection{Effect and Coverage of Semantic Normalization}
\label{sec:rq1}
\begin{figure}[t]
\centering
\includegraphics[width=0.96\columnwidth]{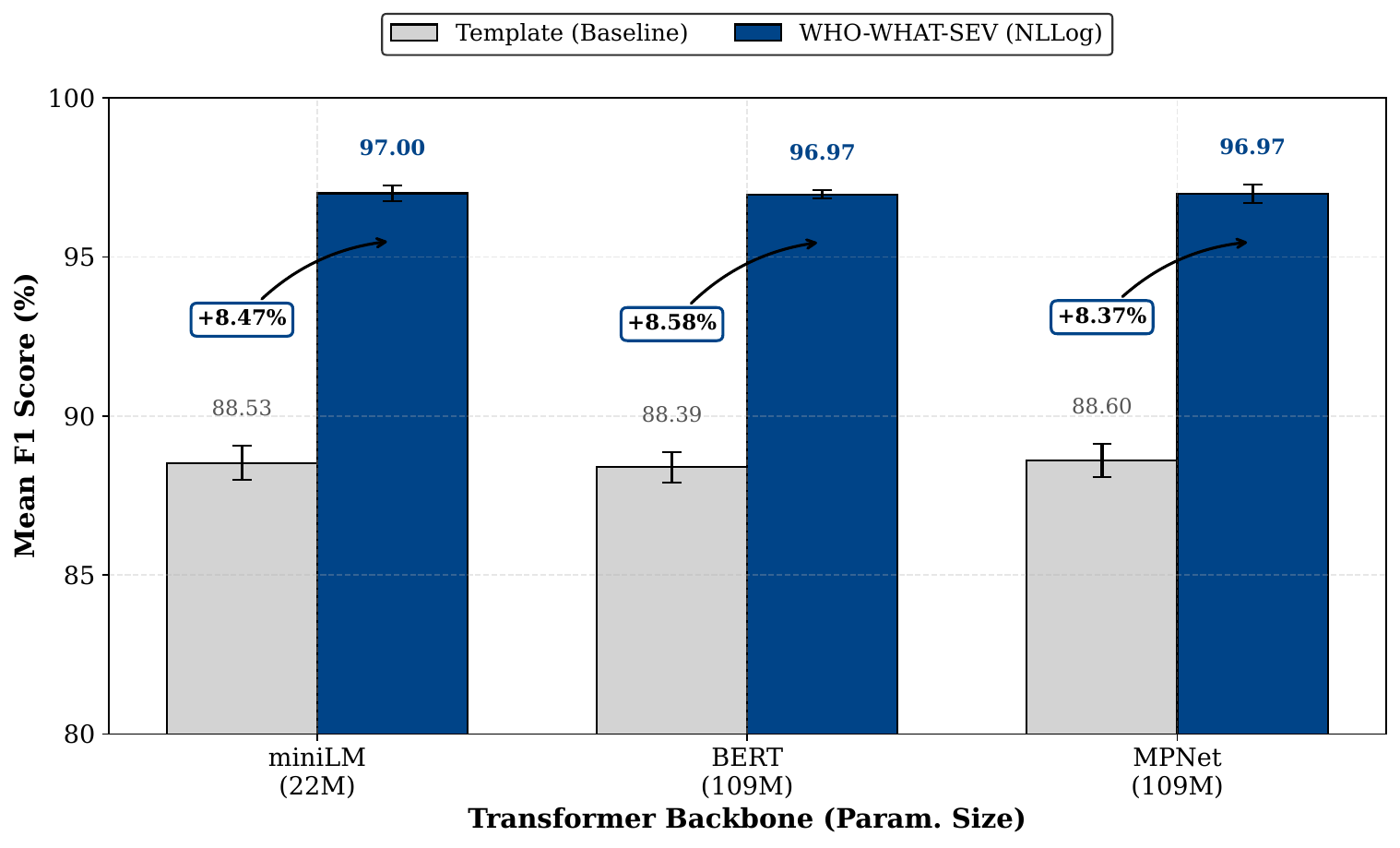}
\caption{Template IDs (T) versus WWS representations across three encoders on BGL. HDFS results are omitted because near-ceiling F\textsubscript{1} ($>99.7\%$ for all variants) compresses the visual comparison.}
\label{fig:representations}
\end{figure}

To assess the effect of semantic normalization, we compare two input formats: raw Drain3 template identifiers (T) and natural-language WWS sentences. Both are encoded with three pre-trained transformers (MiniLM, BERT, and MPNet) and classified with LGBM. The headline comparison is summarized in Table~\ref{tab:rq2_results}: on BGL, WWS normalization lifts mean F\textsubscript{1} from \textbf{88.53\%} (T-MiniLM) to \textbf{97.00\%} (WWS-MiniLM), a gain of 8.47 percentage points. Figure~\ref{fig:representations} extends this comparison across encoders, with less than 0.02\,s additional per-session training time on BGL.

\paragraph{WWS coverage and fallback-only ablation}
To assess portability and isolate the contribution of deterministic lexical refinements, Table~\ref{tab:wws_coverage} characterizes each dataset by the fraction of units served by (a)~general heuristic rules, (b)~optional deterministic lexical refinements, and (c)~the automatic fallback frame, and reports the F\textsubscript{1} of the full WWS pipeline versus a fallback-only variant that suppresses lexical and heuristic refinements where applicable. The ablation separates the effect of full WWS rewriting from the automatic fallback representation, and the answer is corpus-dependent. On BGL, fallback-only WWS preserves performance (97.49\% $\rightarrow$ 97.46\% F\textsubscript{1}), suggesting that the canonicalized BGL templates already expose enough component, severity, and action structure for the frozen encoder. On HDFS, fallback-only WWS reduces F\textsubscript{1} from 99.82\% to 69.50\%, indicating that deterministic lexical refinements carry substantial detection-relevant signal for that corpus. AIT-ADS differs in kind: its WWS analogue is the normalized alert name rather than a template rewrite, so the full and fallback variants coincide by construction. These results make WWS portability measurable rather than automatic. In practice, Table~\ref{tab:wws_coverage} serves as an enrollment-time coverage report: before deploying on a new corpus, operators run the WWS pipeline on training-split logs and inspect the heuristic, lexical, and fallback fractions. For HDFS, 36 of 44 templates required deterministic lexical refinements; for BGL, 95 of 481 templates have such refinements, though the fallback-only ablation shows they are not needed for detection on that corpus. The effort therefore scales with the number of distinct templates, not with log volume.

\begin{table}[htbp]
\centering
\caption{WWS coverage and fallback-only ablation.}
\label{tab:wws_coverage}
\setlength{\tabcolsep}{3.5pt}
\renewcommand{\arraystretch}{1.05}
\resizebox{\columnwidth}{!}{%
\begin{tabular}{lcccccc}
\toprule
\textbf{Dataset} & \textbf{\#Units} & \textbf{Heur.\%} & \textbf{Lex.\%} & \textbf{Fall.\%} & \textbf{Full F\textsubscript{1}(\%)} & \textbf{Fall.\ F\textsubscript{1}(\%)} \\
\midrule
HDFS    & 44  & 18.18  & 81.82 & 0.00  & 99.82\,(0.07) & 69.50\,(1.80) \\
BGL     & 481 & 66.11  & 19.75 & 14.14 & 97.49\,(0.30) & 97.46\,(0.23) \\
AIT-ADS & 92  & 100.00 & 0.00  & 0.00  & 92.75\,(0.69) & 92.75\,(0.69) \\
\bottomrule
\end{tabular}}
\vspace{0.5ex}
{\footnotesize \textit{Note:} \#Units counts Drain3 templates for HDFS and BGL and canonical alert names for AIT-ADS. Heur.\ = general heuristic rules; Lex.\ = deterministic lexical refinements; Fall.\ = automatic fallback (WHO + raw template + SEV). Values are mean\,(std) across five seeds.\par}
\end{table}

\subsection{TF--IDF Representation Weighting}
\label{sec:rq2}
We assess the impact of semantic weighting by comparing three representation strategies: (i) raw template identifiers with mean pooling, (ii) WWS sentence embeddings with uniform mean pooling, and (iii) TF--IDF weighted WWS embeddings (denoted W--WWS). All configurations use MiniLM as the encoder and LGBM as the classifier.

As shown in Table~\ref{tab:rq2_results}, W--WWS yields the best performance among the compared representation modes on both datasets. On HDFS, it achieves the highest F\textsubscript{1} (99.82\%) and the lowest FPR (0.008\%); on BGL, W--WWS raises precision from 82.01\% to 98.22\% and reduces FPR from 2.37\% to 0.197\%, a $12\times$ reduction in false alarms.
\begin{table*}[htbp]
\centering
\caption{Comparison of log representation strategies on HDFS and BGL datasets.}
\label{tab:rq2_results}
\begin{tabular}{llcccc}
\toprule
\textbf{Dataset} & \textbf{Mode} & \textbf{F\textsubscript{1}(\%)} & \textbf{Precision(\%)} & \textbf{Recall(\%)} & \textbf{FPR(\%)} \\
\midrule
\multirow{3}{*}{HDFS} 
  & Template  & 99.76 (0.05) & 99.67 (0.13) & 99.86 (0.07) & 0.010 (0.004) \\
  & WWS       & 99.80 (0.03) & 99.69 (0.12) & \cellcolor{green!20}99.91 (0.09) & 0.009 (0.004) \\
  & W--WWS    & \cellcolor{green!20}99.82 (0.05) & \cellcolor{green!20}99.73 (0.06) & \cellcolor{green!20}99.91 (0.04) & \cellcolor{green!20}0.008 (0.002) \\

\multirow{3}{*}{BGL} 
  & Template  & 88.53 (0.54) & 82.01 (0.96) & \cellcolor{green!20}96.17 (0.27) & 2.37 (0.15) \\
  & WWS       & 97.00 (0.25) & 98.15 (0.74) & 95.88 (0.49) & 0.204 (0.083) \\
  & W--WWS    & \cellcolor{green!20}97.13 (0.26) & \cellcolor{green!20}98.22 (0.76) & 96.06 (0.41) & \cellcolor{green!20}0.197 (0.086) \\
\bottomrule
\end{tabular}
\end{table*}

\paragraph{Sparse text versus dense pipeline}
To isolate dense embedding gains from sparse text alternatives, we add raw-template TF--IDF and WWS-text TF--IDF baselines with logistic regression (LR) under the same 80/20 stratified split and five seeds. Table~\ref{tab:sparse_dense} reports the comparison across the three corpora, and the picture is dataset-dependent. On BGL, replacing the raw-template surface with WWS text alone lifts sparse F\textsubscript{1} from 81.64\% to 91.86\% and cuts FPR from 4.05\% to 0.87\%, and the full dense pipeline reaches 97.49\% F\textsubscript{1} at 0.18\% FPR. On HDFS, however, sparse WWS text \emph{underperforms} sparse raw templates (84.71\% versus 94.33\% F\textsubscript{1}), so some discriminative cues are not preserved by the sparse WWS surface alone; the dense pipeline closes that gap (99.82\% F\textsubscript{1}). On AIT-ADS, the alert-name sparse baseline reaches 91.30\% F\textsubscript{1} (precision 95.55\%, recall 87.43\%, FPR 0.77\%), while the dense pipeline reaches 92.93\% F\textsubscript{1} at the same recall but precision 99.18\% and FPR 0.14\%; the gain is therefore precision- and FPR-driven, not recall-driven. Across the three corpora, the full dense pipeline is consistently strongest.

\begin{table*}[htbp]
\centering
\caption{Sparse text versus dense representation ablation.}
\label{tab:sparse_dense}
\setlength{\tabcolsep}{4pt}
\begin{tabular}{llccccc}
\toprule
\textbf{Dataset} & \textbf{Representation} & \textbf{F\textsubscript{1}(\%)} & \textbf{Precision(\%)} & \textbf{Recall(\%)} & \textbf{FPR(\%)} & \textbf{AUC-PR(\%)} \\
\midrule
\multirow{3}{*}{HDFS}
  & Raw template TF--IDF~+~LR        & 94.33\,(0.28) & 89.35\,(0.52) & 99.91\,(0.05) & 0.36\,(0.02) & 96.05\,(0.25) \\
  & WWS text TF--IDF~+~LR            & 84.71\,(0.24) & 87.46\,(0.45) & 82.13\,(0.12) & 0.36\,(0.01) & 81.30\,(0.20) \\
  & WWS MiniLM~+~TF--IDF~+~LGBM      & \cellcolor{green!15}\textbf{99.82\,(0.07)} & \cellcolor{green!15}\textbf{99.75\,(0.08)} & \cellcolor{green!15}\textbf{99.89\,(0.09)} & \cellcolor{green!15}\textbf{0.01\,(0.00)} & \cellcolor{green!15}\textbf{100.00\,(0.00)} \\
\midrule
\multirow{3}{*}{BGL}
  & Raw template TF--IDF~+~LR        & 81.64\,(0.83) & 72.25\,(0.74) & 93.83\,(1.10) & 4.05\,(0.12) & 91.82\,(0.87) \\
  & WWS text TF--IDF~+~LR            & 91.86\,(0.85) & 92.21\,(0.85) & 91.52\,(1.03) & 0.87\,(0.10) & 97.08\,(0.61) \\
  & WWS MiniLM~+~TF--IDF~+~LGBM      & \cellcolor{green!15}\textbf{97.49\,(0.30)} & \cellcolor{green!15}\textbf{98.37\,(0.56)} & \cellcolor{green!15}\textbf{96.62\,(0.61)} & \cellcolor{green!15}\textbf{0.18\,(0.06)} & \cellcolor{green!15}\textbf{99.70\,(0.07)} \\
\midrule
\multirow{2}{*}{AIT-ADS}
  & Alert-name TF--IDF~+~LR          & 91.30\,(0.66) & 95.55\,(0.68) & 87.43\,(1.01) & 0.77\,(0.12) & 92.39\,(0.52) \\
  & Alert-name MiniLM~+~TF--IDF~+~LGBM & \cellcolor{green!15}\textbf{92.93\,(0.37)} & \cellcolor{green!15}\textbf{99.18\,(0.57)} & 87.43\,(0.68) & \cellcolor{green!15}\textbf{0.14\,(0.10)} & 92.38\,(0.84) \\
\bottomrule
\end{tabular}

\end{table*}

\paragraph{AIT-ADS sparse alert representations}
Table~\ref{tab:ait_sparse_dense} further tests richer non-leaky AIT-ADS sparse alert representations using detector source, event type, and rule-tier (Suricata priority class) tokens. These variants do not improve over alert-name-only sparse TF--IDF, and removing rule-tier has negligible effect, indicating that the AIT-ADS dense result is not caused by an under-specified sparse baseline. We exclude label, attack-stage, timestamp, filename, IP, host, and other high-cardinality identifiers; the AIT-ADS schema has no MITRE tactic or technique fields.

\begin{table*}[htbp]
\centering
\caption{AIT-ADS representation ablation.}
\label{tab:ait_sparse_dense}
\setlength{\tabcolsep}{4pt}
\begin{tabular}{llcccccc}
\toprule
\textbf{Representation} & \textbf{Classifier} & \textbf{F\textsubscript{1}(\%)} & \textbf{Precision(\%)} & \textbf{Recall(\%)} & \textbf{FPR(\%)} & \textbf{AUC-PR(\%)} & \textbf{Train--Test Gap(\%)} \\
\midrule
Alert-name TF--IDF                   & LR   & 91.30\,(0.66) & 95.55\,(0.68) & 87.43\,(1.01) & 0.77\,(0.12) & 92.39\,(0.52) & 0.23\,(0.85) \\
Rich alert TF--IDF                   & LR   & 90.80\,(0.27) & 94.68\,(0.26) & 87.23\,(0.63) & 0.92\,(0.05) & 92.40\,(0.66) & 0.21\,(0.60) \\
Rich alert TF--IDF, no severity      & LR   & 90.94\,(0.56) & 95.03\,(0.42) & 87.18\,(0.82) & 0.86\,(0.07) & 92.41\,(0.62) & 0.19\,(0.76) \\
Alert-name MiniLM~+~TF--IDF          & LGBM & \cellcolor{green!15}\textbf{92.93\,(0.37)} & \cellcolor{green!15}\textbf{99.18\,(0.57)} & 87.43\,(0.68) & \cellcolor{green!15}\textbf{0.14\,(0.10)} & 92.38\,(0.84) & 0.44\,(0.51) \\
\bottomrule
\end{tabular}
\end{table*}

\subsection{Encoder Size and Efficiency}
\label{sec:rq1h3}
\begin{figure*}[htbp]
\centering
    \includegraphics[width=1.5\columnwidth]{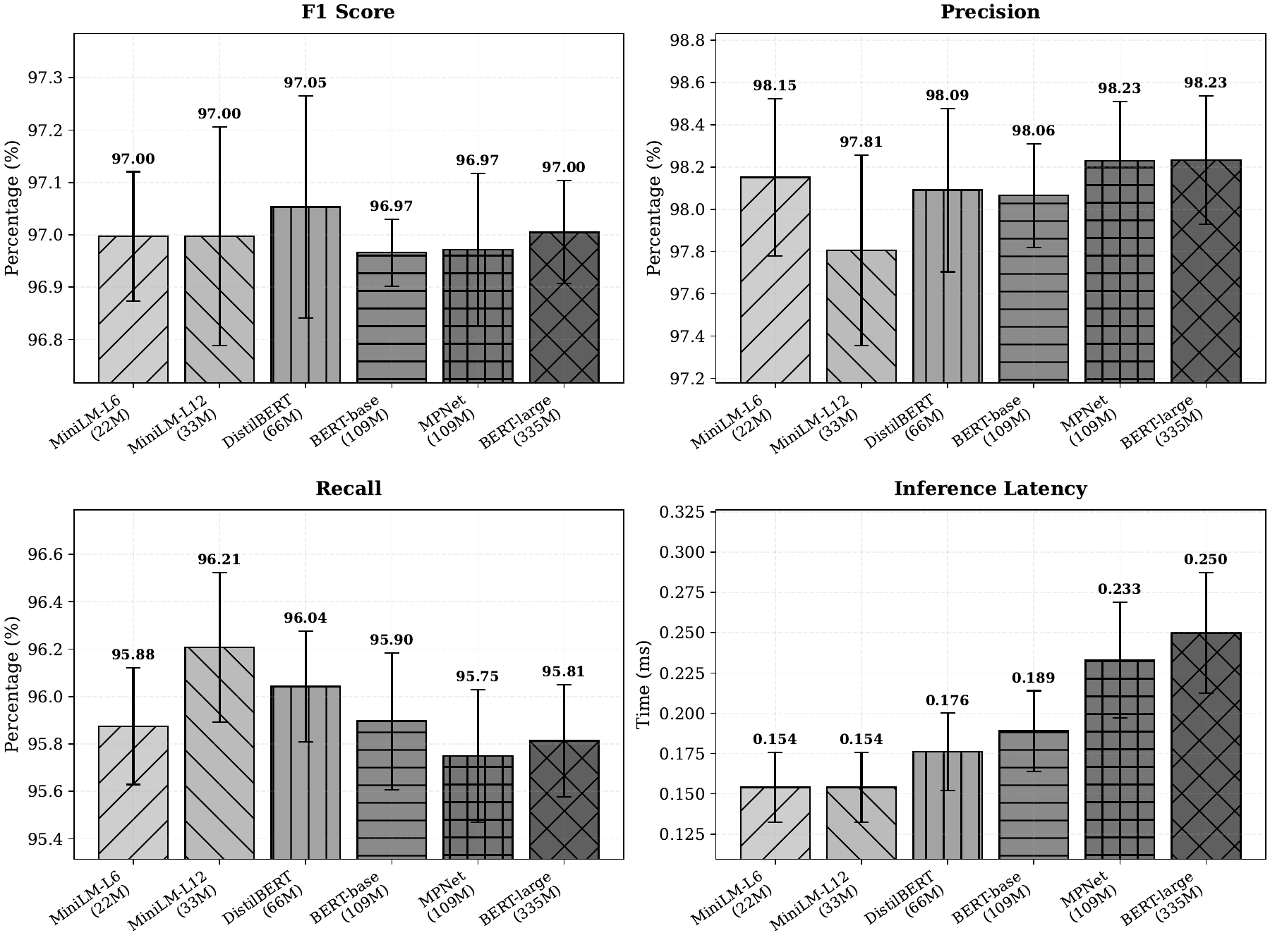}
    \caption{Encoder accuracy--latency trade-off on BGL.}
    \label{fig:encodersize}
\end{figure*}

Fig.~\ref{fig:encodersize} compares six frozen encoders on BGL with W--WWS inputs. All models reach F\textsubscript{1} $\geq 96.9\%$; MiniLM-L6 is the most favorable accuracy-latency point, matching the larger encoders at roughly 60\% lower per-sentence latency (consistent with Table~\ref{tab:latency-breakdown}).

\subsection{Comparison Against Baselines}
\label{sec:rq1h4}
We compare NLLog against two reproduced baselines under the matched HDFS/BGL protocol (DeepLog and LogBERT) and report published results from four additional systems (LogAnomaly, LogLLM, LogGPT, and LogLLaMA) for context. Results are summarized in Table~\ref{tab:model_performance}.

On \textbf{BGL}, NLLog reaches 97.67\% F\textsubscript{1} with the highest precision among the matched-protocol methods (99.46\%); this is 11.6~pp above reproduced DeepLog and 6.8~pp above reproduced LogBERT. On \textbf{HDFS}, NLLog attains 99.81\% F\textsubscript{1} with 99.62\% precision and 100.00\% recall, again improving over both reproduced baselines.
\begin{table*}[htbp]
\centering
\caption{Performance comparison of anomaly detectors.}
\label{tab:model_performance}
\begin{tabular}{lcccccc}
\toprule
\multirow{2}{*}{Model} & \multicolumn{3}{c}{BGL} & \multicolumn{3}{c}{HDFS} \\
 & {Prec.(\%)} & {Rec.(\%)} & {F\textsubscript{1}(\%)} & {Prec.(\%)} & {Rec.(\%)} & {F\textsubscript{1}(\%)} \\
\midrule
\multicolumn{7}{l}{\textit{Reproduced under matched protocol}} \\
DeepLog~\cite{du2017deeplog}     & 89.74 & 82.78 & 86.12 & 88.44 & 69.49 & 77.34 \\
LogBERT~\cite{guo2021logbert}    & 89.40 & 92.32 & 90.83 & 87.02 & 78.10 & 82.32 \\
\textbf{NLLog \emph{[Ours]}} & \cellcolor{green!15}\textbf{99.46} & 95.94 & \cellcolor{green!15}\textbf{97.67} & \cellcolor{green!15}\textbf{99.62} & \cellcolor{green!15}\textbf{100.00} & \cellcolor{green!15}\textbf{99.81} \\
\midrule
\multicolumn{7}{l}{\textit{Published results (different protocols, included for context)}} \\
LogAnomaly~\cite{meng2019loganomaly} & 97.00 & 94.00 & 96.00 & 96.00 & 94.00 & 95.00 \\
LogLLM~\cite{guan2024logllm}      & 86.10 & 97.90 & 91.60 & 99.40 & 100.00 & 99.70 \\
LogLLaMA~\cite{yang2025logllama}    & 92.75 & 99.33 & 95.93 & 93.90 & 85.25 & 89.36 \\
LogGPT~\cite{10386543}      & 94.00 & 97.70 & 95.80 & 88.40 & 92.10 & 90.10 \\
\bottomrule
\end{tabular}

\vspace{0.5ex}
  {\footnotesize
  \textit{Note:} Tables~\ref{tab:rq2_results}--\ref{tab:idf_poisoning} use different fixed, ablation, budget-tuned, and stress-test LGBM configurations; values should be compared within each table.\par}
\end{table*}

\subsection{Effect of Classifier Choice}
\label{sec:rq2h5}
We evaluated nine classifiers spanning classical machine learning (ML) and deep learning (DL) using W--WWS embeddings on BGL, HDFS, and AIT-ADS. Ensemble tree models (LGBM, XGB, RF) consistently deliver the highest F\textsubscript{1} across all datasets, with low FPRs and sub-millisecond inference latency; Appendix~\ref{app:classifiers} (Fig.~\ref{fig:rq2_classifiers}) shows the full accuracy--latency trade-off. Deep models (multilayer perceptron, MLP; deep neural network, DNN) approach tree-based performance on HDFS and BGL but degrade on AIT-ADS, while support vector machines (SVM), logistic regression (LR), convolutional neural networks (CNN), and long short-term memory networks (LSTM) underperform in this pooled-embedding setting.
\subsection{Data-Fraction Robustness}
\label{sec:rq2h6}

We measured F\textsubscript{1} as a function of training size on BGL (20\%, 40\%, 60\%, 80\%) using NLLog's best configuration (WWS + TF--IDF + MiniLM) against three reproduced baselines: DeepLog (RNN-based), DeepCASE~\cite{vanede2022deepcase} (attention-based), and LogBERT (transformer-based).

\begin{figure}[htbp]
    \centering
    \includegraphics[width=\linewidth]{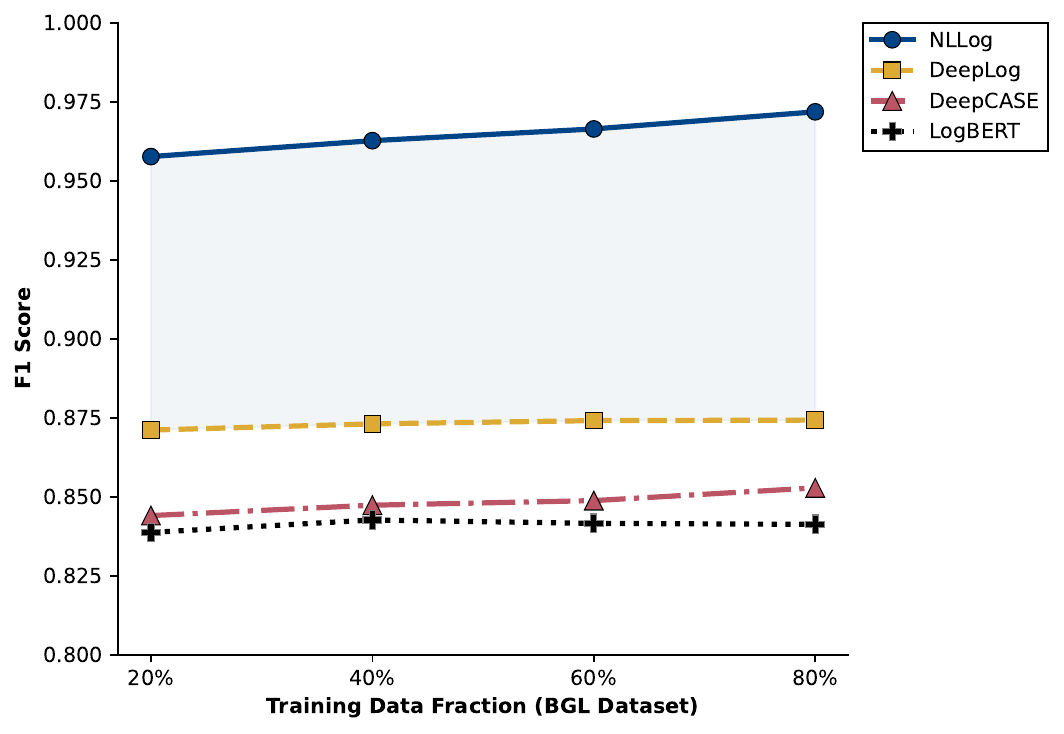}
    \caption{BGL F\textsubscript{1} versus training-data fraction.}
    \label{fig:rq2_generalization}
\end{figure}

Figure~\ref{fig:rq2_generalization} shows that under our matched protocol NLLog achieves higher F\textsubscript{1} than the three reproduced baselines at every training fraction. At 20\% training data, NLLog reaches 95.9\% F\textsubscript{1}, ahead of reproduced DeepLog (87.2\%), DeepCASE (84.3\%), and LogBERT (83.9\%); performance rises steadily to 97.3\% at 80\%, while the baselines plateau earlier.
\subsection{Detection under Alert Budgets}
\label{sec:rq3-h7}
SOC operations often impose strict alert-volume limits, often framed as false-positive budgets such as $\leq$2\% FPR~\cite{sharma2021survey,axelsson2000base}. We therefore evaluate NLLog under a 2\% FPR cap using five classifiers on BGL. Appendix~\ref{app:alert_budget_curves} plots the ROC and precision-recall curves.
Table~\ref{tab:rq5-bgl} quantifies the results. At the selected LGBM operating point, NLLog reaches 97.67\% F\textsubscript{1}, 99.46\% precision, 95.94\% recall, and 0.06\% FPR; under the FPR\,$\leq$\,2\% constraint, LGBM achieves 99.69\% recall. XGB and RF follow closely, whereas SVM and LR produce higher FPRs.
\begin{table*}[htbp]
\centering
\caption{Performance on BGL under an FPR $\leq$ 2\% alert budget.}
\label{tab:rq5-bgl}
\begin{tabular}{lcccccc}
\toprule
Classifier & F\textsubscript{1}(\%) & Precision(\%) & Recall(\%) & Recall@FPR$\leq$2\% & FPR(\%) & AUC-PR(\%) \\
\midrule
LGBM           & \cellcolor{green}{97.67} & \cellcolor{green}{99.46} & \cellcolor{green}{95.94} & \cellcolor{green}{99.69} & \cellcolor{green}{0.06} & \cellcolor{green}{99.71} \\
XGB       & 97.12 & 99.45 & 94.90 & 99.66 & \cellcolor{green}{0.06} & 99.66 \\
RF       & 95.96 & 99.33 & 92.81 & 99.60 & 0.07 & 99.60 \\
SVM                & 94.80 & 98.76 & 91.15 & 99.20 & 0.13 & 99.20 \\
LR & 94.27 & 96.10 & 92.50 & 98.51 & 0.42 & 98.52 \\
\bottomrule
\end{tabular}
\end{table*}

\subsection{Interpretability and Analyst-Facing Evidence}
\label{sec:rq3_h8}

NLLog augments the LGBM classifier with (i) exact TreeSHAP log-odds attributions $\boldsymbol{\phi}_j$ (Eq.~\eqref{eq:treeshap-logit}) and (ii) a responsibility matrix that projects those attributions to individual sentences, yielding $\psi_{ji}$ and template-level scores $\Psi_{j,t}$ (Eqs.~\eqref{eq:psi-ji}--\eqref{eq:template-agg}). The analyst-facing artifact is a ranked top-$k$ list ($k \le 15$) of WWS sentences with signed contribution scores, formatted as a compact alert-evidence record. Each row uses \textcolor{red}{$\uparrow$} for an anomaly-promoting and \textcolor{blue}{$\downarrow$} for a normality-promoting contribution, with the bracketed SHAP value $\phi_i$ and the template frequency ``$\times k$''; from the base value $\phi_0=-14.76$, a positive cumulative total yields an anomalous prediction. Listings~\ref{lst:shap_tn} and~\ref{lst:shap_tp} show condensed HDFS examples; Appendix~\ref{app:shap_examples} (Fig.~\ref{fig:rq3_local}) gives the matching waterfall views, including a contrasting false positive.

\begin{lstlisting}[
    basicstyle=\scriptsize\ttfamily,
    caption={HDFS true-negative SHAP explanation (top-4 by $|\phi|$).},
    label={lst:shap_tn}
]
"session_idx": 0, "label": 0, "pred": 0, "prob": 1.5e-07, "category": "TN"
↓[-0.1649]  datanode data receiver: received block <block_id> from <ip><port> destined for <ip><port> (info) ×3
↓[-0.0943]  datanode packet responder: received block <block_id> of size <num> from <ip> (info) ×3
↓[-0.0885]  datanode packet responder: <*> for block <block_id> <status> (info) ×3
↓[-0.0720]  filesystem namespace manager: delete <block_id> is added to invalid set of <ip> <port> (info) ×3
\end{lstlisting}

\begin{lstlisting}[
    basicstyle=\scriptsize\ttfamily,
    caption={HDFS true-positive SHAP explanation (top-4 by $|\phi|$).},
    label={lst:shap_tp}
]
"session_idx": 22, "label": 1, "pred": 1, "prob": 0.9999998, "category": "TP"
↑[+2.1169]  filesystem namespace manager: delete <block_id> is added to invalid set of <ip> <port> (info) ×4
↑[+1.8128]  filesystem dataset manager: deleting block <block_id> file <*> (info) ×4
↑[+1.5827]  filesystem namespace manager: block map updated <ip> <port> is added to <block_id> size <num> (info) ×4
↑[+1.1394]  filesystem namespace manager: received a redundant addstoredblock request for block <block_id> on <ip><port> size <*> (warning) ×1
\end{lstlisting}

In the TN, routine DataNode packet-handling templates dominate with downward contributions; in the TP, namespace- and dataset-manager deletion templates contribute large positive $\phi_i$, pushing the log-odds well above zero ($p\approx0.9999998$).

\begin{figure}[t]
\centering
\includegraphics[width=0.96\linewidth]{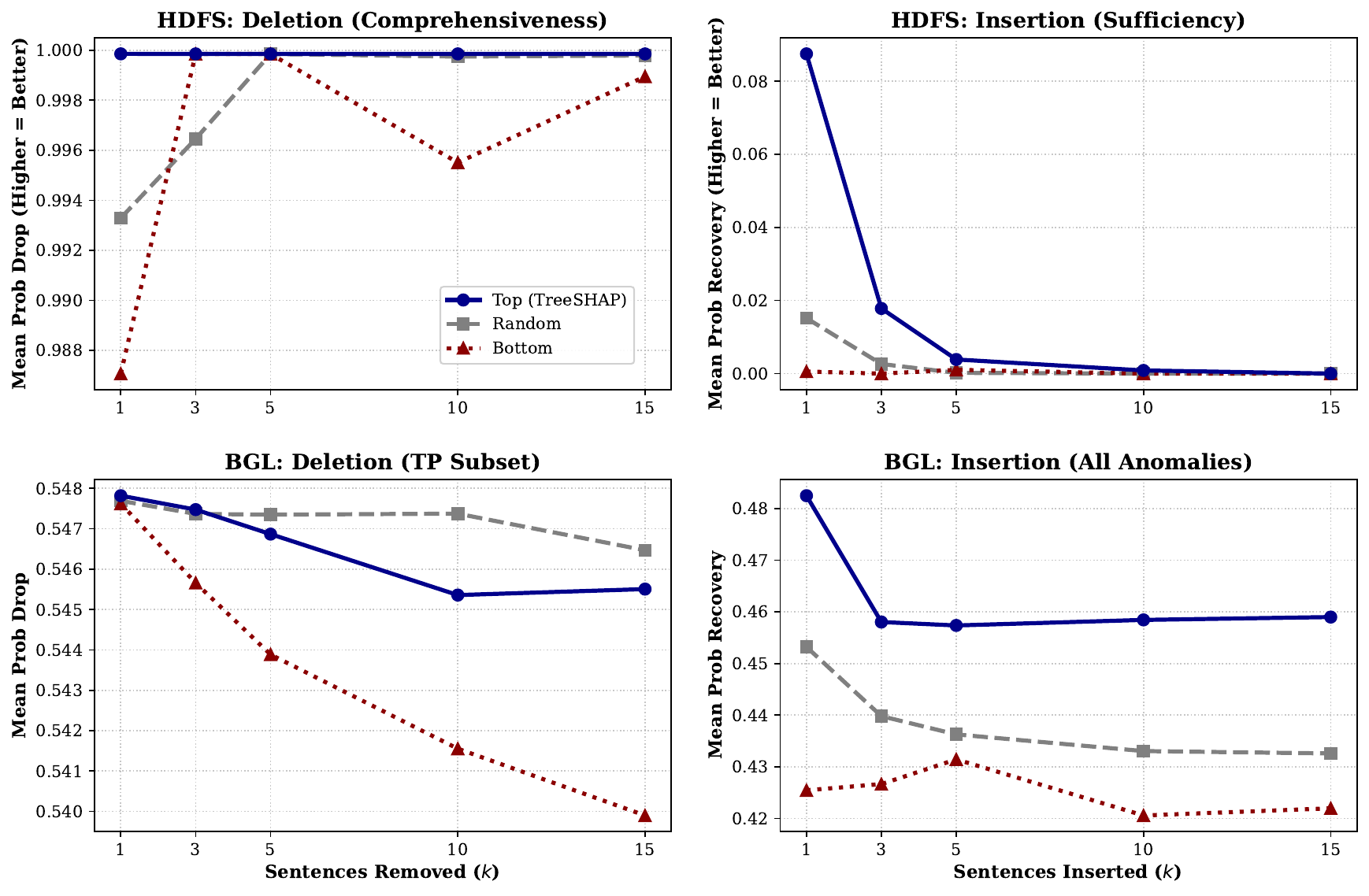}
\caption{Faithfulness tests for TreeSHAP sentence attribution.}
\label{fig:faithfulness}
\end{figure}

\paragraph{Faithfulness of sentence attribution}
We assess decision faithfulness of the TreeSHAP sentence-attribution layer using deletion, insertion, and sufficiency tests with random and bottom-$k$ controls on anomalous test sessions. On BGL, we report deletion-based area over the perturbation curve (AOPC) on the true-positive subset because anomalous sessions that are already false negatives under the base classifier have near-zero baseline anomaly probability and can destabilize normalized deletion scores. On this subset, removing TreeSHAP top-ranked sentences yields a mean normalized drop of 0.552 and an unnormalized drop of 0.547. The top-versus-random insertion gap is strictly positive at every $k \in \{1,3,5,10,15\}$, with a 95\% paired-bootstrap confidence interval excluding zero at $k=1$ (gap 0.029, CI $[0.013, 0.046]$). On HDFS, all anomalous test sessions are true positives, both AOPC forms are effectively identical at 1.000, and the $k=1$ insertion gap is 0.072 with CI $[0.066, 0.079]$. Figure~\ref{fig:faithfulness} shows that top-ranked sentences consistently dominate random and bottom-ranked controls, supporting the claim that the back-projected TreeSHAP scores capture decision-relevant evidence on HDFS and BGL; analogous faithfulness tests for AIT-ADS remain future work.

\subsection{Adversarial Evaluation}
\label{sec:adv-eval}
\begin{figure}[htbp]
    \centering
    \includegraphics[width=\linewidth]{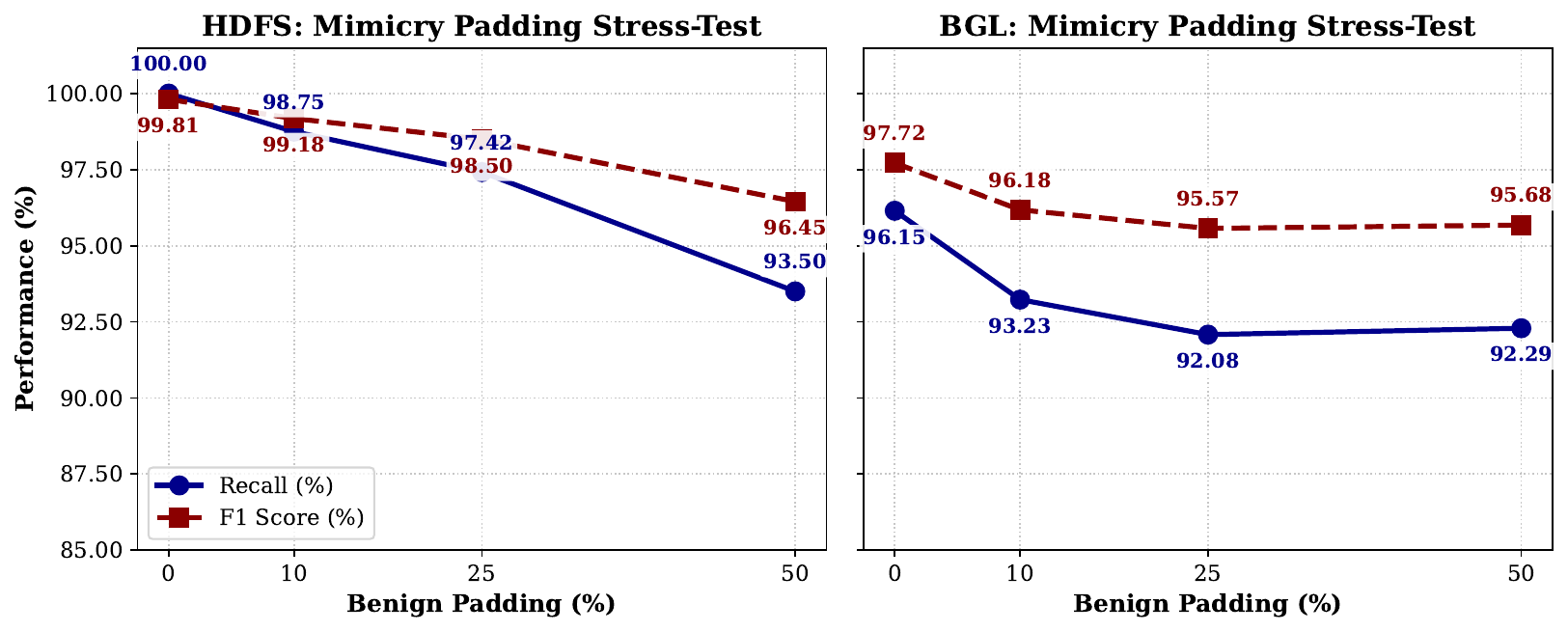}
    \caption{Mimicry padding sweep on HDFS and BGL.}
    \label{fig:mimicry_padding}
\end{figure}

To stress-test detection and the compact top-$k$ evidence window under benign-noise dilution, we introduce \emph{mimicry padding} and \emph{top-$k$ hit-rate} analyses.
Mimicry padding inserts $p\%$ benign WWS lines into anomalous sessions, simulating a lightweight adversary that dilutes anomaly signals without modifying templates. As shown in Fig.~\ref{fig:mimicry_padding}, HDFS recall drops from 100\% to 93.5\% at 50\% padding while FPR remains negligible ($\approx 0.01\%$), indicating that dilution can reduce recall without materially changing false-positive behavior. BGL shows the same qualitative pattern more mildly: recall stays above 92\% and F\textsubscript{1} above 95\% even at 50\% padding.
To evaluate explanation robustness, we compute Hit@$k$: the fraction of anomalous sessions where at least one anomaly-labeled or anomaly-associated line appears in the top-$k$ SHAP-ranked lines (BGL uses corpus line-level labels; HDFS uses anomaly-associated templates as a proxy). Anomalies manifest as \emph{distributed patterns} across correlated log lines rather than single sentinel events, with the first anomaly-associated line at median rank~8; the ranking is therefore a \emph{compact evidence window} rather than a pinpoint localization mechanism. Table~\ref{tab:topk_stability} shows that Hit@15 stays at or above 90\% and the median first-anomaly rank at or below 9 across the tested padding levels; Hit@1 and Hit@5 are near-zero for the same distributional reason (see table footnote).
To separate test-time dilution from retraining-time corpus-statistics contamination, Appendix~\ref{app:idf_poisoning} (Table~\ref{tab:idf_poisoning}) reports an IDF-poisoning stress test in which only the TF--IDF fit corpus is contaminated while held-out session text remains unchanged. Headline metrics degrade only slightly (HDFS F\textsubscript{1}: 99.81\%\,$\rightarrow$\,99.79\%; BGL: 97.72\%\,$\rightarrow$\,97.51\%), but representation drift and probability shifts remain measurable, supporting the use of frozen clean IDF statistics during deployment.

\begin{table}[htbp]
    \centering
    \caption{Top-$k$ explanation stability under mimicry padding (BGL).}
    \label{tab:topk_stability}
    \begin{tabular}{rcc}
        \toprule
        Padding (\%) & Hit@15 (\%) & Median Rank \\
        \midrule
        0  & 100.00 & 8.00 \\
        10 & 100.00 & 8.00 \\
        25 &  90.00 & 8.00 \\
        50 &  90.00 & 9.00 \\
        \bottomrule
    \end{tabular}

    \vspace{0.5ex}
    {\footnotesize \textit{Note:} Hit@1 is 0.00\% at every padding level; Hit@5 is 0.00\% at 0\% and 10\% padding and 10.00\% at 25\% and 50\% padding. Both stay near zero because anomalies appear as correlated line clusters (median first-anomaly rank~8) rather than as single sentinel events.\par}
\end{table}
\section{Related Work}
\label{sec:related_work}

Log anomaly detection has developed along several distinct lines. Classical approaches emphasize parsing, template extraction, and invariant mining~\cite{he2017drain,lin2016logcluster,he2016experience,lou2010invariantmining,zhu2022survey,meng2020survey}. These methods are often interpretable and efficient, but they typically operate on event IDs, counts, or mined rules rather than semantically richer text, and they may require substantial feature engineering or retuning as log formats evolve.

Sequence-oriented deep models such as DeepLog and LogAnomaly~\cite{du2017deeplog,meng2019loganomaly} model temporal regularities over template streams, while semantic models such as PLELog, NeuralLog, and LogBERT~\cite{yang2021plelog,le2021neurallog,guo2021logbert} introduce transformer-style representations into the pipeline. More recent systems including LogLLaMA~\cite{yang2025logllama}, LogPrompt~\cite{10554918}, LogGPT~\cite{10386543}, and LogLLM~\cite{guan2024logllm} explore prompt-based or decoder-based formulations. NLLog is closest to this semantic family, but it makes a different systems trade-off: rather than relying on raw-log language modeling or heavier decoder-style inference, it first applies deterministic template-to-language normalization and then uses lightweight pooled representations, tree ensembles, and TreeSHAP attribution. We therefore emphasize the end-to-end normalization, pooling, and attribution pipeline rather than encoder scale.

Interpretability for log analysis remains comparatively underdeveloped. Attention-based or contextual systems such as DeepCASE and DeepEAD~\cite{vanede2022deepcase,wang2023deepead} offer analyst-facing cues, while broader interpretable ML work~\cite{chen2019looks,koh2017influence} provides general techniques for tracing model behavior. NLLog returns ranked WWS sentences from a pooled session representation via TreeSHAP and an explicit responsibility projection, framed as anomaly attribution rather than causal reconstruction.

Provenance-graph systems such as ORTHRUS, PROGRAPHER, and ThreatRace~\cite{orthrus,prographer,threatrace} target richer audit telemetry and system-wide attack tracing, often with graph construction overhead, heavier instrumentation, and GPU-backed training; they are useful context but not protocol-matched baselines for sessionized line logs. Recent empirical critiques also show that log-anomaly results can depend strongly on dataset construction and evaluation protocol~\cite{le2022log,ali2025ladstudy}, which motivates our matched-protocol comparisons and explicit separation of reproduced baselines from published reference points.
\section{Discussion}
\label{sec:discussion}

We close with a synthesis of the design choices that drive NLLog's behavior, the limitations and threats to validity revealed by the evaluation, and the operational posture under which the system should be deployed.

\subsection{Synthesis of Key Findings}

Two design choices drive most of NLLog's behavior. First, deterministic WWS rewriting supplies the analyst-readable surface form returned as evidence, with corpus-dependent fallback sufficiency characterized by the enrollment-time coverage report (Section~\ref{sec:rq1}, Table~\ref{tab:wws_coverage}). Second, dense MiniLM embeddings with TF--IDF pooling and LGBM consistently outperform sparse TF--IDF alternatives (Table~\ref{tab:sparse_dense}), cutting FPR by an order of magnitude or more on HDFS and BGL and raising AIT-ADS precision while lowering FPR at matched recall. Together, the auditable deterministic rewrite combined with lightweight dense encoding provides a measurable representation layer for SOC triage whose coverage assumptions are surfaced before deployment.
\subsection{Limitations and Threats to Validity}
\paragraph{Scope and Threat Assumptions}
NLLog does not defend against adversaries who can modify log contents, inject fabricated templates, or suppress logging; such attacks violate the trusted-logging assumption and require complementary log-integrity or provenance mechanisms. The sequence-only and cross-host attack classes excluded by NLLog's design are stated in Section~\ref{sec:threat}. Analyst utility and time savings remain to be measured in a controlled user study. Mitigations beyond our current scope include template integrity attestation, cross-host corroboration, and freezing IDF statistics from a clean enrollment period.

\paragraph{Evasion by Dilution}
Because session vectors are TF--IDF weighted averages, the architecture is inherently vulnerable to \emph{flooding attacks}. Section~\ref{sec:adv-eval} and Appendix~\ref{app:idf_poisoning} quantify this under mimicry padding and IDF poisoning: test-time mimicry padding is the stronger operational threat, while retraining-time corpus contamination is mitigated by frozen lexical statistics from a clean enrollment period.

\paragraph{Evaluation scope and refinement effort}
Chronological and sessionization evaluations on AIT-ADS and faithfulness tests on HDFS/BGL provide deployment-oriented checks; analogous temporal splits for HDFS and BGL, broader sessionization sweeps, and AIT-ADS faithfulness remain future work. The fallback-only ablation also limits claims of zero-effort portability; the enrollment-time coverage check discussed next surfaces this requirement. Appendix~\ref{app:limits} (Table~\ref{tab:limits}) summarizes these and other limitations with planned mitigations.
\subsection{Portability and Operational Posture}
Deployments should run the enrollment-time coverage report described in Section~\ref{sec:rq1} before relying on fallback behavior. The WWS layer is fail-soft because no event is dropped, but fallback-only rewriting is not always sufficient. Overfitting risk is limited by design: WWS rewriting is deterministic, TF--IDF weighting uses unlabeled corpus statistics, and only the final classifier is learned (Appendix~\ref{app:hyperparams}); all stages scale linearly in the number of log events. Our evaluation runs on a 32-core Xeon Gold 6242 with no GPU and under 2\,GB of peak memory, with Drain3 and TF--IDF in streaming mode and outputs emitted as JavaScript Object Notation (JSON) records compatible with standard SIEM alert pipelines. Live deployment validation remains future work.

\section{Ethical Considerations and Artifact Availability}
\label{sec:ethics}

NLLog is evaluated on public benchmark datasets and does not involve human-subject experimentation. Canonicalization masks high-cardinality tokens such as IPs, ports, and identifiers, which reduces direct exposure of volatile or potentially sensitive values in downstream representations. Deployment errors nevertheless remain possible: false positives can increase analyst workload, and false negatives can defer investigation of real incidents. We therefore position NLLog as a triage aid rather than an autonomous decision maker, and recommend analyst oversight for any operational use.

NLLog is intentionally self-contained: local encoders, no external API calls, and commodity CPUs suffice for inference. Artifact-release details appear in Appendix~\ref{app:artifact}. NLLog itself does not use external LLM services during detection; editorial LLM use is disclosed in the LLM Usage Statement at the end of the paper.

\section{Conclusion}
\label{sec:conclusion}

We introduced \textbf{NLLog}, a log-to-language pipeline that deterministically rewrites parsed templates into analyst-readable WWS sentences, embeds them with a frozen encoder, pools them with TF--IDF, and back-projects tree-ensemble decisions to ranked evidence. Across HDFS, BGL, and AIT-ADS, NLLog achieves high F\textsubscript{1} and low false-positive rates at roughly 10\,ms median CPU latency, with chronological, sessionization, and faithfulness checks supporting the headline results. Fallback and sparse ablations identify when generic rewriting is sufficient and when corpus-specific deterministic refinement is needed, yielding a focused systems claim: deterministic, language-aligned rewriting provides an auditable representation layer for lightweight SOC anomaly triage.

\bibliographystyle{IEEEtran}
\bibliography{mybibliography}

\appendices
\section{Canonicalization and Template Determination}
\label{app:canonicalization}\label{app:drain3}
NLLog applies a fixed sequence of regular-expression substitutions, following prior log-analysis practice~\cite{he2016experience,he2017drain}, that replaces volatile tokens with canonical placeholders such as \texttt{<IP>}, \texttt{<PORT>}, \texttt{<BLOCK>}, \texttt{<NUM>}, and \texttt{<EXC>}. This preserves the message skeleton while removing high-cardinality noise and scrubbing potentially sensitive values.

We use Drain3 with depth $d=4$ and \texttt{max\_children}=100. Canonicalized logs are routed through the fixed-depth parse tree, exact matches are preferred over wildcards, and a new cluster is created only when similarity to an existing template falls below the configured threshold. Each cluster stores its \texttt{event\_id}, frequency, and representative template.
\section{Classifier Comparison Details}
\label{app:classifiers}
\begin{figure*}
    \centering
    \includegraphics[width=\linewidth]{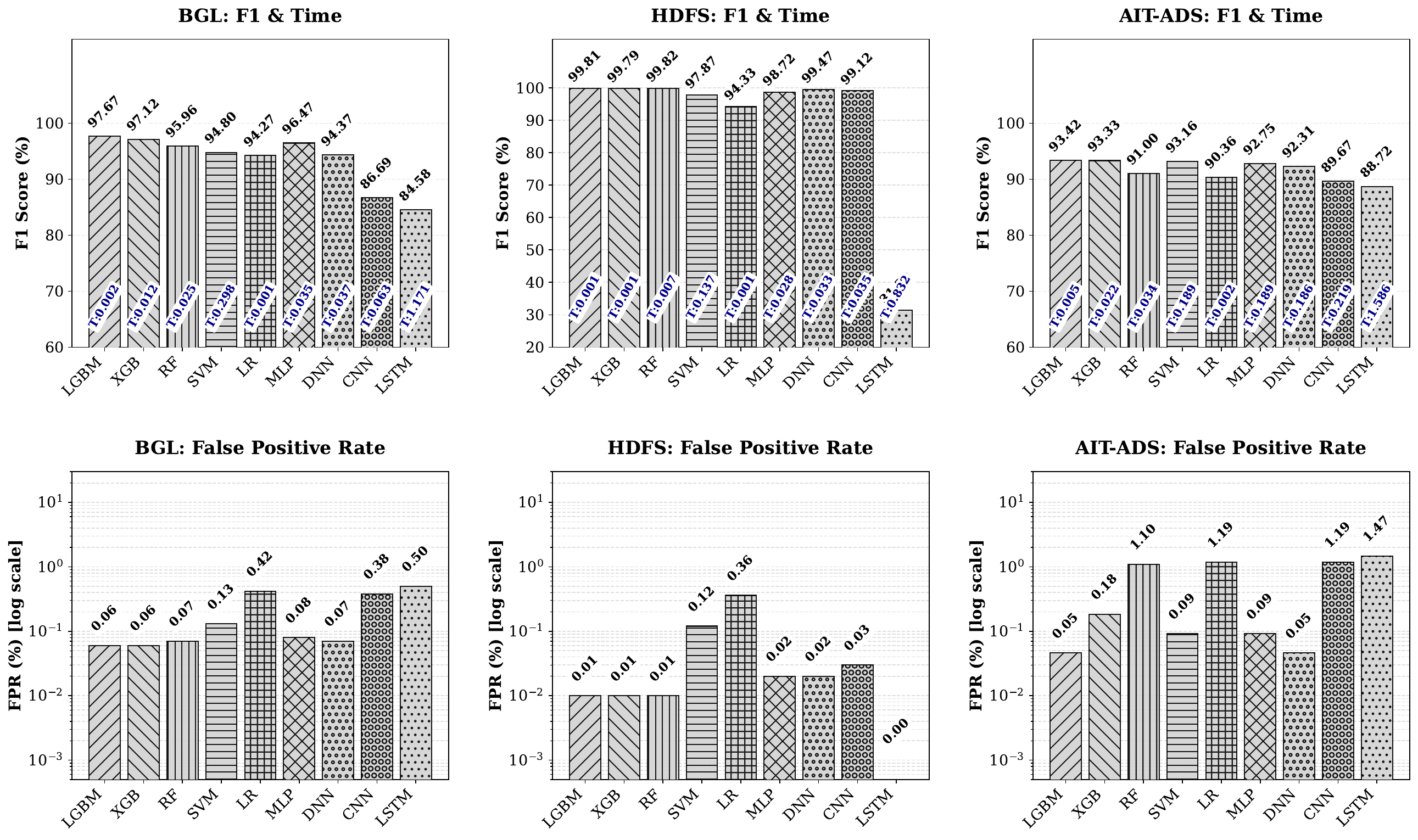}
    \caption{Classifier evaluation: F\textsubscript{1} with per-session inference time (top) and false-positive rate on log scale (bottom) across the three datasets.}
    \label{fig:rq2_classifiers}
\end{figure*}

Figure~\ref{fig:rq2_classifiers} reports F\textsubscript{1} with inference time (top row) and false-positive rate (bottom row, log scale) for the nine classifiers compared in Section~\ref{sec:rq2h5}.
\section{Alert-Budget Curves}
\label{app:alert_budget_curves}

\begin{figure}[H]
\centering
\includegraphics[width=\linewidth]{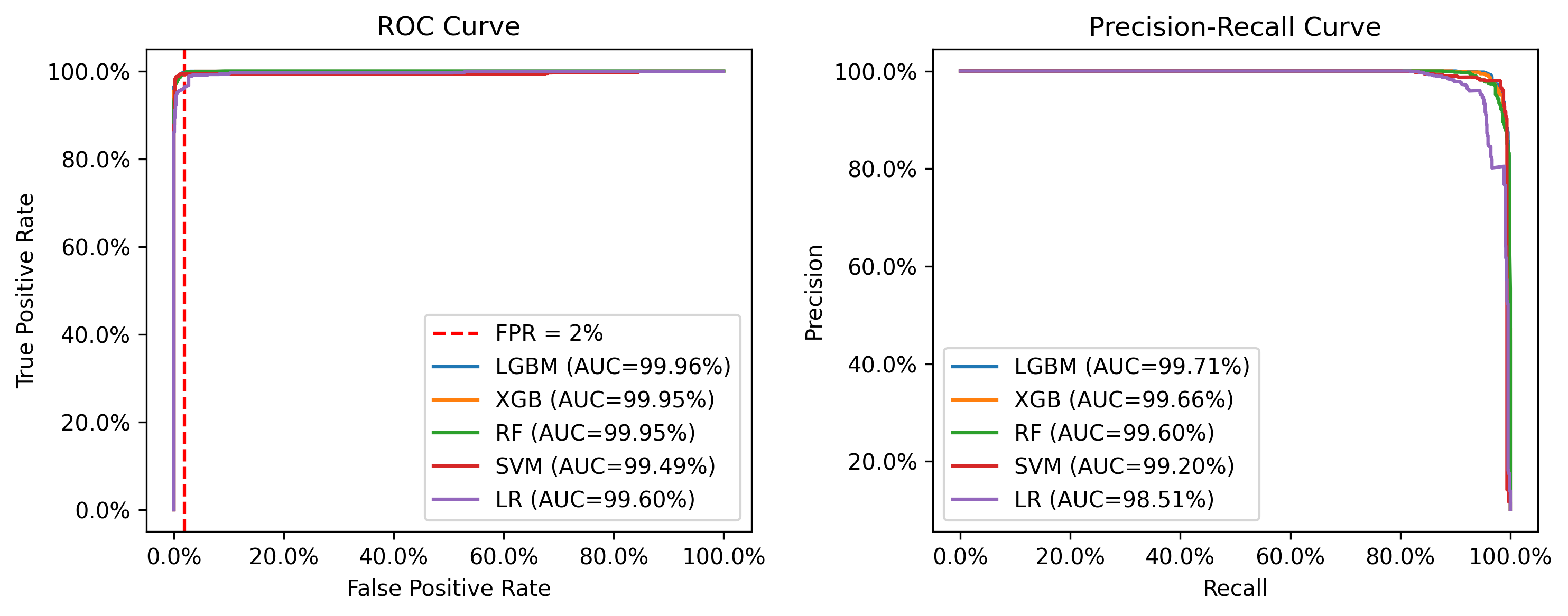}
\caption{ROC and precision-recall curves on BGL under the FPR\,$\leq$\,2\% alert budget.}
\label{fig:rq5_rocpr}
\end{figure}

Figure~\ref{fig:rq5_rocpr} accompanies the BGL alert-budget analysis in Section~\ref{sec:rq3-h7}.
\section{IDF-Poisoning Stress Test}
\label{app:idf_poisoning}
Table~\ref{tab:idf_poisoning} reports the full IDF-poisoning numbers behind the summary in Section~\ref{sec:adv-eval}, including the clean baseline, contaminated IDF, and frozen-clean-IDF settings on HDFS and BGL.

\begin{table*}[htbp]
\centering
\caption{IDF-poisoning stress test on HDFS and BGL.}
\label{tab:idf_poisoning}
\begin{tabular}{llcccc}
\toprule
\textbf{Dataset} & \textbf{Setting} & \textbf{Prec. (\%)} & \textbf{Rec. (\%)} & \textbf{F\textsubscript{1} (\%)} & \textbf{FPR (\%)} \\
\midrule
\multirow{3}{*}{HDFS}
& Clean train / clean test & 99.62 & 100.00 & 99.81 & 0.01 \\
& Contaminated IDF         & 99.62 & 99.97  & 99.79 & 0.01 \\
& Frozen clean IDF         & 99.62 & 100.00 & 99.81 & 0.01 \\
\midrule
\multirow{3}{*}{BGL}
& Clean train / clean test & 99.35 & 96.15 & 97.72 & 0.07 \\
& Contaminated IDF         & 99.03 & 96.04 & 97.51 & 0.11 \\
& Frozen clean IDF         & 99.35 & 96.15 & 97.72 & 0.07 \\
\bottomrule
\end{tabular}

\vspace{0.5ex}
{\footnotesize \textit{Note:} Mean L2 session drift under contamination is 0.002280 (HDFS) and 0.008316 (BGL); maximum absolute probability change is 0.781591 (HDFS) and 0.606948 (BGL).\par}
\end{table*}

\section{AIT-ADS Sessionization Sensitivity}
\label{app:aitads_window}
\begin{figure}[H]
\centering
\includegraphics[width=\linewidth]{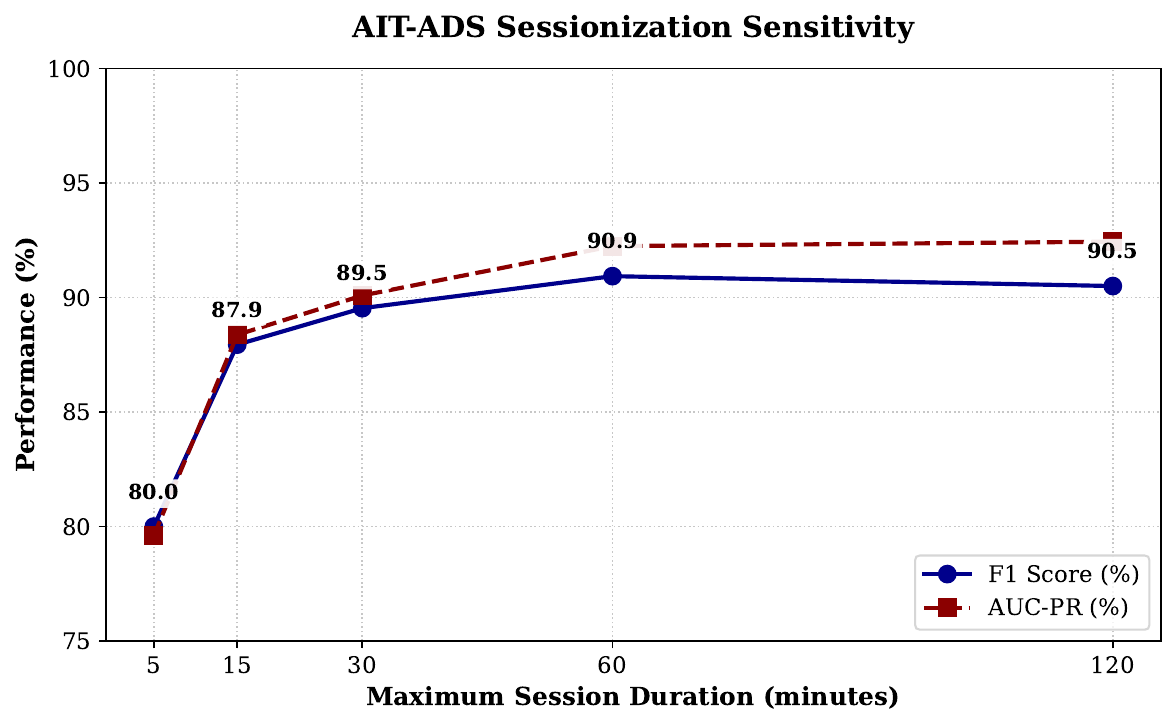}
\caption{AIT-ADS sessionization sensitivity across maximum session durations with a fixed 300\,s idle timeout.}
\label{fig:aitads_window_sensitivity}
\end{figure}
Figure~\ref{fig:aitads_window_sensitivity} plots the AIT-ADS sessionization sweep summarized in Section~\ref{sec:exp-setup}: test F\textsubscript{1} and AUC-PR across maximum session durations of 5, 15, 30, 60, and 120 minutes with the idle timeout fixed at 300\,s.

\section{Waterfall Views of Sentence Attribution}
\label{app:shap_examples}
\begin{figure*}[htbp]
  \centering
  \subfloat[FP sample session]{\includegraphics[width=0.75\textwidth]{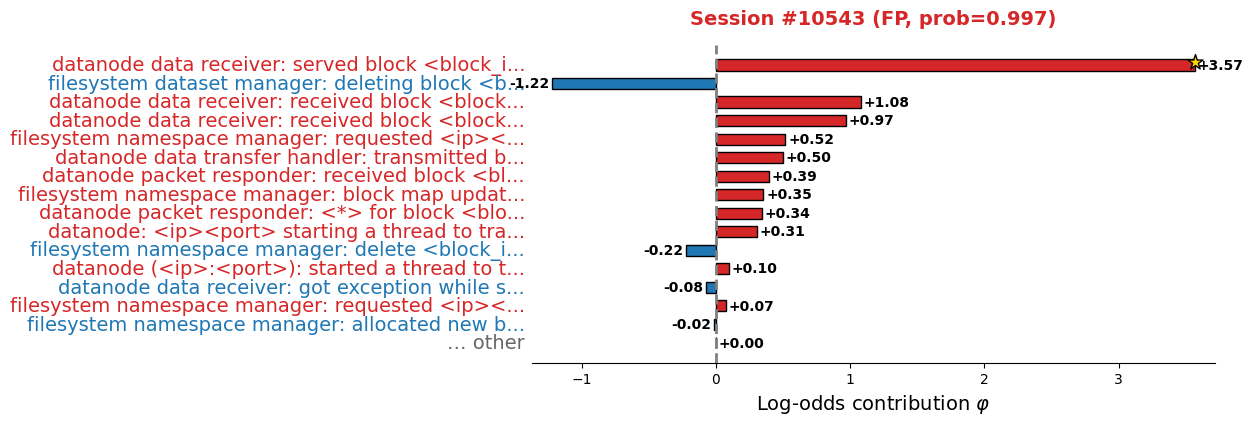}}
  \hfill
  \subfloat[TP sample session]{\includegraphics[width=0.75\textwidth]{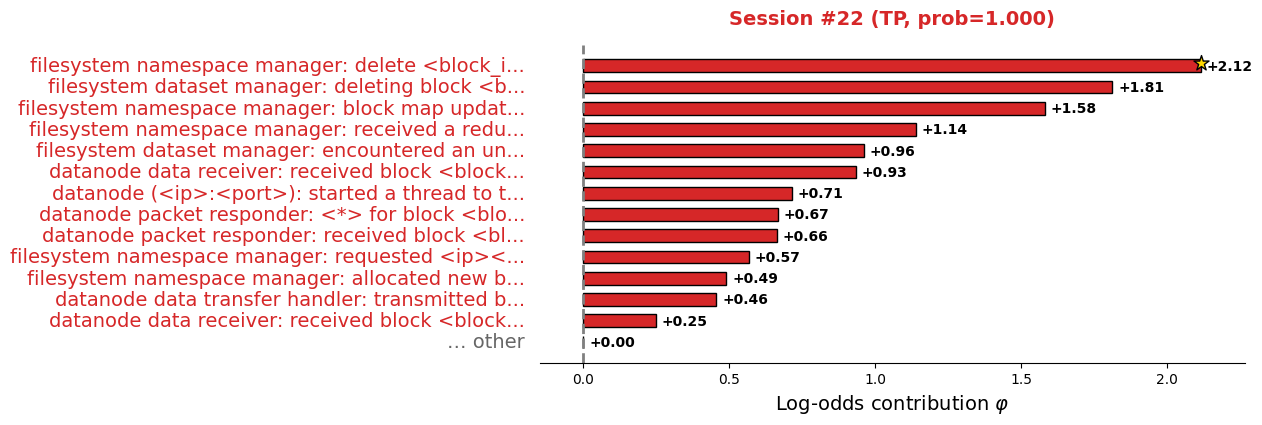}}
  \caption{HDFS sentence-level log-odds attribution for a false positive and a true positive. Positive bars increase the anomaly score; negative bars decrease it.}
  \label{fig:rq3_local}
\end{figure*}

Figure~\ref{fig:rq3_local} provides graphical waterfall views complementing Listings~\ref{lst:shap_tn} and~\ref{lst:shap_tp} in Section~\ref{sec:rq3_h8}, contrasting the same HDFS true-positive session with a false positive (\#10543, $\hat{p}=0.997$); each waterfall shows at most 15 sentences. Full per-session JSON explanations are included in the artifact package.

\section{NLLog Limitations}
\label{app:limits}
Table~\ref{tab:limits} summarizes the main limitations of NLLog together with corresponding mitigations or future work directions.
\begin{table*}[htbp]
  \caption{Key limitations and planned mitigations.}
  \label{tab:limits}
  \centering
  \begin{tabularx}{0.99\textwidth}{>{\raggedright\arraybackslash}p{0.1\textwidth}XX}
    \toprule
    {Dimension} & {Limitation} & {Mitigation / Future Work} \\
    \midrule
    Template fidelity & Unseen or attacker-injected templates may lack deterministic lexical refinements & Drain3 flags them as \texttt{<UNKNOWN>} (surfaced via SHAP); harden with signed binaries and cross-source corroboration; explore deterministic rule induction or audited template-to-WWS suggestion tools \\
    Loss of rare value signals & Canonicalization masks outlier IPs/IDs            & Combine value-frequency baselines with semantic scores \\
    Dataset bias             & Public corpora may not reflect proprietary logs     & Field deployments and continuous-learning studies \\
    Streaming drift          & Temporal robustness is only evaluated on AIT-ADS    & Extend chronological splits and drift studies to HDFS/BGL; explore incremental TF--IDF and sliding-window SHAP caching \\
    Human factors            & Analyst utility not measured in a controlled user study & Controlled SOC usability studies \\
    \bottomrule
  \end{tabularx}

  \vspace{0.5ex}
\end{table*}
\FloatBarrier
\section{Hyper-parameters and Artifact Availability}
\label{app:hyperparams}\label{app:artifact}
Complete JSON dumps of selected non-default hyperparameters are included in the reproducibility package under \texttt{artifacts/hparams/}. If accepted, we will submit an anonymized artifact package for ACSAC artifact evaluation, including code, configuration files, canonicalization rules, WWS mappings, hyperparameter files, and scripts needed to reproduce the reported tables and figures. For third-party datasets, we will provide acquisition and preprocessing instructions rather than redistributing material whose license prohibits repackaging. Any optional repository provided during review will be anonymized to preserve the dual-anonymous submission process.

\FloatBarrier
\section*{LLM Usage Statement}
LLMs were used for editorial purposes in this manuscript, and all outputs were inspected by the authors to ensure accuracy and originality. No LLM was used to generate experimental results, numerical findings, or scientific claims.

\end{document}